 \documentclass[final,authoryear,3p,times]{elsarticle}

\usepackage{amsmath,amssymb}
\usepackage{epsfig,subfigure,color}

\newcommand{\vecv}{\ensuremath{\boldsymbol{v}}}
\newcommand{\vecg}{\ensuremath{\boldsymbol{g}}}
\newcommand{\vecnab}{\ensuremath{\boldsymbol{\nabla}}}
\newcommand{\DD}{\ensuremath{\mathrm{D}}}
\newcommand{\dd}{\ensuremath{\mathrm{d}}}

\journal{Astronomy and Computing}

\begin{document}

\begin{frontmatter}



\title{\textcolor{black}{Influence of adaptive mesh refinement and the hydro solver
	on shear-induced mass stripping in a minor-merger scenario}}


\author[ws]{W. Schmidt\corref{cor1}}
\ead{schmidt@astro.physik.uni-goettingen.de}

\author[js]{J. Schulz}
\ead{schulz@math.uni-goettingen.de}

\author[li]{L. Iapichino}
\ead{luigi.iapichino@lrz.de}

\author[fv]{F. Vazza}
\ead{franco.vazza@hs.uni-hamburg.de}

\author[aa]{A. S. Almgren}
\ead{asalmgren@lbl.gov}

\cortext[cor1]{Corresponding author}

\address[ws]{Institut f\"ur Astrophysik, Georg-August-Universit\"at G\"ottingen, Friedrich-Hund Platz 1, D-37077 G\"ottingen, Germany}
\address[js]{Institut f\"ur Numerische und Angewandte Mathematik, Georg-August-Universit\"at G\"ottingen, Lotzestra{\ss}e 16-18, D-37083 G\"ottingen, Germany}
\address[li]{Leibniz-Rechenzentrum der Bayerischen Akademie der Wissenschaften, Boltzmannstr.~1, D-85748 Garching b.~M\"unchen, and
	Universit\"at Heidelberg, Zentrum f\"ur Astronomie, Institut f\"ur Theoretische Astrophysik, Philosophenweg 12, D-69120 Heidelberg, 
	Germany}
\address[fv]{Hamburger Sternwarte, Gojenbergsweg 112, D-21029 Hamburg, Germany}
\address[aa]{Center for Computational Sciences and Engineering, Lawrence Berkeley National Laboratory, Berkeley,
CA 94720, USA}

\begin{abstract}
\textcolor{black}{We compare two different codes for simulations of cosmological structure formation to investigate
the sensitivity of hydrodynamical instabilities to numerics, in particular, the hydro solver and the application of
adaptive mesh refinement (AMR). As a simple test problem, we consider an initially spherical gas cloud in a wind, 
which is an idealized model for the merger of a subcluster or galaxy with a big cluster.
Based on an entropy criterion, we calculate the mass stripping from the subcluster as a function of time.
Moreover, the turbulent velocity field is analyzed with a multi-scale filtering technique.
We find remarkable differences between the commonly used PPM 
solver with directional splitting in the {\sc Enzo} code and an unsplit variant of PPM in the {\sc Nyx} code, 
which demonstrates that different codes can converge to systematically different solutions even when using uniform grids.
For the test case of an unbound cloud, AMR simulations reproduce uniform-grid results for the mass stripping quite well, 
although the flow realizations can differ substantially. If the cloud is bound by a static gravitational potential, 
however, we find strong sensitivity to spurious fluctuations which are induced at the cutoff radius of the potential
and amplified by the bow shock.
This gives rise to substantial deviations between uniform-grid and AMR runs performed with {\sc Enzo},
while the mass stripping in {\sc Nyx} simulations of the subcluster is nearly independent of 
numerical resolution and AMR. Although many factors related to numerics are involved, our study indicates that
unsplit solvers with advanced flux limiters help to reduce grid effects and to keep numerical noise under control,
which is important for hydrodynamical instabilities and turbulent flows. 
}
\end{abstract}


\end{frontmatter}


\section{Introduction and astrophysical motivation}
\label{sc:intro}

In the widely accepted paradigm of hierarchical formation of cosmic structure, large virialized objects like clusters of galaxies grow by accretion of smaller subclusters. In this process, which proceeds up to the current epoch, the merging of halos and subhalos is believed to be an important source of turbulence in the intra-cluster medium \citep[hereafter ICM; e.~g.,][]{PaulIapi11,VazzaBru11} together with other mechanisms like the baroclinic injection of vorticity at curved shocks \citep{krc07,ib12}, outflows of active galactic nuclei \citep{hby06,ss06a,bsh09} and the motion of single galaxies through the ICM \citep{k07,ro11}. The importance of cluster mergers goes beyond being mere stirring agents in the ICM. For example, they are strongly correlated with the occurrence of central cluster diffuse radio emission (radio halos; \citealt{ceg10}) via some still debated mechanism of cosmic ray acceleration \citep{bj14}. Moreover, major mergers launch shock waves in the ICM, which are observed as brightness and temperature edges in X-ray images \citep{m10}. There is also the prospect of measuring turbulence with the up-coming Astro-H mission \citep{BiffiDol12,ShangOh13}.

The role of mergers as injectors of bulk flow and turbulence in the ICM has been recognized in hydrodynamical simulations of the build-up of galaxy clusters \citep[e.~g.,][]{r98,nb99,in08,VazzaBru11,VazzaRoed12,m14}. Complementary to cosmological simulations starting from realistic initial conditions, idealised cluster simulations are useful to study mergers in a simplified setup and with a better control on the problem parameters \citep{rbl96,rs01,mbb07}.
A special subclass of cluster mergers is constituted by minor mergers, where one the merging objects has a much smaller mass than the other. Although these processes do not have the same impact on the energy budget of the ICM as major mergers, they are interesting in their own right. For example, they are associated with observed structures like merger cold fronts \citep{mv07,RussFab14}. During minor mergers, the interface between the host ICM and the subcluster is subject to the Kelvin-Helmholtz instability, which results in gas stripping and injection of turbulence in the downstream region 
\citep{ssh06,RussFab14,EckMol14}. 
This process has been noticed in full cosmological simulations \citep{in08,mis09} and has been studied in greater detail by means of idealised setups, following the infall of a low-mass subcluster 
or an elliptical galaxy into the static ICM of a big cluster as it is seen from an observer moving with the subcluster \citep{hcf03,t05a,xcd07,afm07,dp07,RoedKr14,RoedKrII}.

The study of \citet{IapiAda08} belongs to the latter class of simulations. 
A particularly interesting aspect of this study is the transition from laminar flow to turbulence 
in the boundary layer of the subcluster, which is difficult to tackle for compressible hydro solvers with numerical viscosity. 
In this article, we further elaborate on this computational problem by comparing simulations carried out with the cosmological
fluid dynamics codes {\sc Enzo} and {\sc Nyx}, which implement a split and an unsplit variant of the widely applied 
piecewise parabolic method (PPM, \citealt{ColWood84}).
In order to gain a clearer insight into the simulations performed with the setup from \citet{IapiAda08}, we also address the simpler case
of an unbound cloud in a wind as in \citet{AgerMoore07}. In this case, the cloud is initially defined as a spherical region 
of higher gas density in pressure equilibrium with the ambient medium.
The code comparison by \citet{AgerMoore07} was a seminal work that demonstrated striking differences between 
smoothed particle hydrodynamics (SPH) and grid-based codes.

Another question concerns an assumption that is often taken for granted in computational astrophysics, 
namely the equivalence between a run performed with a uniform grid and the corresponding simulation using adaptive mesh refinement (AMR)
at the same effective spatial resolution. Recently, \citet{m14} has questioned that dynamic refinement reproduces 
turbulent fluid properties, particularly if the refinement method is based on keeping the mass in a cell roughly constant. 
Thus, we want to investigate in a systematic way under which conditions statistical agreement between computations with AMR
and uniform grids at the same effective resolutions can be achieved. 
By computing statistics related to the stripping of mass from the subcluster and by investigating the flow structure, 
a significant impact of refinement strategies in AMR simulations has been shown by \citet{IapiAda08}.
In particular, AMR based on local gradients of density or temperature is not able to follow the 
formation of the turbulent subcluster wake, whereas this is possible with criteria based on the variability of structural invariants of 
the flow. This means that thresholds for refinement are calculated from statistical moments of scalars such as the squared vorticity 
(cf.~\citealt{SchmFeder09,Schmidt}).
 
To infer the impact of the different hydro solvers implemented in {\sc Enzo} and in {\sc Nyx} and to compare
uniform-grid versus AMR runs, we compute the mass stripped from the subcluster as a function of time.
For the definition of the cloud mass, we propose a criterion that is based on an entropy threshold. 
We find systematic differences, which are further analyzed by means of 
the multi-scale filtering approach of \citet{VazzaRoed12}. After explaining our methodology in Section~\ref{sc:methods}
in more detail, the results for the simple cloud  without gravity and the subcluster are presented in Sections~\ref{sc:blob} and \ref{sc:msc}, respectively. Our conclusions are presented in the last section.
 
\section{Numerical methods and simulations}
\label{sc:methods}

We consider both gravitationally bound and unbound variants of the cloud in a wind, as defined by
\citet{IapiAda08}. As initial condition, we set a spherically symmetric isothermal cloud in
pressure equilibrium with a homogeneous background medium with temperature 
$k_{\rm B}T_{\rm b} = 8.0\;\mathrm{keV}$ and density  $\rho_{\rm b}=7.9\times 10^{-28}\;\mathrm{g\,cm^{-3}}$. 
In the simple case of an unbound cloud, we assume a sphere of radius $250\;\mathrm{kpc}$ with
constant density $\rho_{\rm c}=6.3\times 10^{-27}\;\mathrm{g\,cm^{-3}}$. The condition
of pressure equilibrium implies a temperature $k_{\rm B}T_{\rm c}=1.0\;\mathrm{keV}$ inside the cloud.
To produce a wind in $x$-direction, an inflowing boundary condition with a uniform
velocity $v_{\rm b}=1.6\times 10^3\,\mathrm{km\,s^{-1}}$ is imposed at the left face of the domain.
The boundary conditions at the other faces of the domain are outflowing. 
Since the cloud is not anchored by a gravitational well, it drifts in the downstream direction. For this
reason, we use an elongated box of size $16\times4\times 4\;\mathrm{Mpc}$.
Apart from the chosen scales, this setup is similar to the blob test of \citet{AgerMoore07}.

\citet{IapiAda08}, on the other hand, assume that the cloud is bound by an external gravitational
potential, which corresponds to a static dark-matter halo with a King profile:
\begin{equation}
	\label{eq:dm-profile}
	\rho_\mathrm{dm}(r) = \rho_\mathrm{dm,c} \left[1 + \left(
    	\frac{r}{r_{\rm core}} \right)^2 \right]^{-3 / 2}\,. 
\end{equation}
The initial density profile of the cloud is obtained by integrating the equation of hydrostatic
equilibrium for the central density $\rho_{\rm c}=0.1\rho_\mathrm{dm,c}=6.3\times 10^{-27}\,\mathrm{g\,cm^{-3}}$,
constant temperature $k_{\rm B}T_{\rm b} = 3.65\;\mathrm{keV}$, and the core radius $r_{\rm core}=250\,\mathrm{kpc}$.
We use the same setup here, except for a cutoff at the radius $r_{\rm max}=6r_{\rm core}$. 
The Cartesian coordinates of the cloud centre ($r=0$) are $(0.4,0.5,0.5)\times 4\;\mathrm{Mpc}$ in a cubic domain of
$4\;\mathrm{Mpc}$ linear size. For $r>r_{\rm max}$,
the gravitational acceleration is set to zero and the state is given by the state of the background medium.
The cutoff is necessary because of the applied boundary conditions, which are the same as in the case without gravity.
In principle, the wind could be assumed to be in a turbulent state. 
However, the properties of turbulence in the ICM are quite uncertain and there is no suitable method that
would allow us to self-consistently add turbulence to the background medium. Artificially added perturbations 
at the inflow boundary would largely decay before the could reach the cloud. Consequently, we consider only turbulence 
that is produced by hydrodynamical instabilities in this study. This has furthermore the advantage that 
perturbations of numerical origin can be investigated in a clear manner. For brevity,
we subsequently refer to the cloud with static gravitational potential as ``subcluster".

To compute the gas-dynamical evolution for an adiabatic equation of state with $\gamma=5/3$, we apply the 
cosmological AMR codes {\sc Enzo} (ascl:1010.072; \citealt{SheaBry04,Enzo13})\footnote{While we used version 2.3
	for the unbound cloud problem, the subcluster simulations were computed with the older version 2.1.
	We repeated selected runs with version 2.3, but did not find substantial differences 
	that would affect the conclusions drawn in this article.} and {\sc Nyx} \citep{AlmBell13}. 
As described in the method paper by \citet{Enzo13}, {\sc Enzo} uses directionally split PPM
\citep{ColWood84}. A variety of Riemann solvers are available 
in the current code version. We are using the default two-shock approximation \citep[see][]{Toro97}, with 
the Harten-Lax-van Leer scheme as fallback. We do not make use of the dual energy formalism, as the flow 
in the problem under consideration does not have Mach numbers much higher than unity.
The {\sc Nyx} code, as described in \citet{AlmBell13}, features an
unsplit version of PPM with full corner coupling \citep{MillCol02}. The Riemann solver in
{\sc Nyx} also uses a two-shock approximation.  Differences between {\sc Enzo}  and {\sc Nyx}
include the split vs unsplit nature of the integrator, as well as the
details of the reference state used in the PPM profile, the flux limiters,
and the exact discretization of the gravitational source terms in the
energy equation. In addition, the construction of refined grid patches in
both codes follows the same clustering algorithm as described in \citet{BergRig92} 
but {\sc Nyx}, unlike {\sc Enzo}, does not require that each child grid have
a unique parent. Moreover, there are different control parameters to constrain the
size of grid patches as well as the buffer zones between consecutive
refinement levels. Finally, the solution procedure for the multilevel
gravitational field differs between the two codes.
Both codes feature N-body solvers to compute the evolution of dark matter, but we use only a
static gravitational potential. Self-gravity of the gas is not included either because the halo is the main 
source of gravity. Also other physical processes such as viscous damping, cooling, star formation, and magnetic fields 
are not considered here.

As shown by \citet{IapiAda08}, simple refinement by overdensity does not fully capture
the frontal bow shock and the turbulent wake produced by vortex shedding from the cloud. 
For this reason, we apply refinement by the vorticity modulus and 
the rate of compression \citep{SchmFeder09,Schmidt}. 
The vorticity is the curl of the velocity, $\boldsymbol{\omega}=\vecnab\times\vecv$,
and the rate of compression is the negative substantial time derivative of the divergence $d=\vecnab\cdot\vecv$. 
To compute the rate of compression, we evaluate the source terms in the dynamical equation for 
$d$, which follows by contracting the partial differential equation for the velocity 
with the divergence operator \citep[see also][]{IapiSchm11,SchmColl13}:
\begin{equation}
  \label{eq:div}
  -\frac{\DD d}{\DD t}
  = \frac{1}{2}\left(|S|^{2}-\omega^{2}\right) + \vecnab\cdot\left(\frac{\vecnab P}{\rho}\right) -\vecnab\cdot\vecg_{\rm K}.
\end{equation}
The first term on the right-hand side describes the competing effects of strain and vorticity, the following term accounts for the effect of the thermal gas pressure, and the last term corresponds to the gravity resulting 
from the King profile (see equation~\ref{eq:dm-profile}).
For $q_1=\omega^2$ and $q_2=-\DD d/\DD t$, cells are tagged for refinement to the next higher level if
\begin{equation}
	|q_i| \ge |\langle q_i\rangle_n| + \max\left(|\langle q_i\rangle_n|,\mathrm{std}_n q_i\right),
\end{equation} 
where the brackets $\langle \cdot\rangle_n$ denote the average over all cells at level $n$ and 
$\mathrm{std}_n q_i=\langle q_i^2\rangle_n-\langle q_i\rangle_n^2$ is the standard deviation from the average. The basic rationale
of this method is that refinement should be triggered if the local value exceeds the typical fluctuation given by the standard
deviation, \textcolor{black}{but only if the fluctuation is significant, i.~e., comparable to the mean or larger.
Thus, we define the thresholds for refinement by level-wise statistics of 
the control variables vorticity and rate of compression, without any tunable parameters.}

\begin{table}[t]
  \begin{center}
      \begin{tabular}{rclr}
        \hline
        $N_0$ & $n_{\rm max}$ & AMR & $\Delta_{\rm min}\;[\mathrm{kpc}]$ \\
       \hline\hline            
       \multicolumn{4}{c}{unbound cloud}\\
       \hline
       $512\times 128^2$  & $0$ & no & $31.3$ \\
       $512\times 128^2$  & $1$ & yes & $15.6$ \\
       $1024\times 256^2$ & $0$ & no & $15.6$ \\
       $512\times 128^2$  & $2$ & yes & $7.8$ \\
       $2048\times 512^2$ & $0$ & no & $7.8$ \\
       \hline                
       \multicolumn{4}{c}{subcluster}\\
       \hline
       $128^3$  & $1$ & yes & $15.6$ \\
       $256^3$  & $0$ & no & $15.6$ \\
       $128^3$  & $2$ & yes & $7.8$ \\
       $512^3$  & $0$ & no & $7.8$ \\
       $128^3$  & $3$ & yes & $3.9$ \\
       \hline                
      \end{tabular}
  \end{center}
  \caption{Overview of simulation runs, ordered by increasing numerical resolution, where 
  	$N_0$ is the number of root-grid cells, $n_{\rm max}$ the maximum refinement level,
	and $\Delta_{\rm min}$ the cell size at highest refinement level.}
  \label{tb:runs}
\end{table}

We performed the simulations listed in Table~\ref{tb:runs}, using both {\sc Enzo} and {\sc Nyx}. 
The smallest resolved length scale in the uniform-grid
simulations with $2048\times 512^2$ (unbound cloud) or $512^3$ (subcluster) grid cells
is $\Delta_{\rm min}\approx 7.8\;\mathrm{kpc}$.
These simulations can be compared to AMR simulations with the same effective resolution,
which is reached with root-grid resolutions
of $512\times 128^2$ (unbound cloud) or $128^3$ (subcluster) at the second level of refinement. 
Since the domain volume for a given resolution is four times larger 
for the unbound cloud than for the subcluster, three levels of refinement 
($\Delta_{\rm min}\approx 3.9\;\mathrm{kpc}$) were applied only in the case of the subcluster. 

In order to better highlight the small-scale motions in the downstream of the stripped cloud, 
we employ a {\it multi-scale} filtering technique which makes no assumptions on the injection scale of chaotic motions
\citep{VazzaRoed12}. 
The filter reconstructs the local mean velocity field around each cell by iteratively computing the 
local mean velocity field as
\begin{equation} 
	\overline{\vecv(L_n)} =\frac{\sum_{i}(r<L_n)w_{\rm i}\vecv_i} {\sum_{i} w_i},
\end{equation} 
where $w_{\rm i}$ is a weighting function (e.~g., gas density or gas mass, but for our simulations we simply set $w_{\rm i}=1$ 
because of the small density contrast). The local small-scale velocity field is computed as 
$\delta \vecv(L_{\rm n})=\vecv-\overline{\vecv(L_n)}$ for increasing values of $L_n$. The iterations are continued until the
relative variation of the local turbulent velocity between two iterations is below the given tolerance parameter, which based on 
our tests we set to $\epsilon=0.1$. Additionally, the scheme stops iterating whenever the skewness of the velocity distribution 
with a stencil of $8^3$ cells is found to be larger than a tolerance parameter, as specified in  \citet{VazzaRoed12}. As final output, 
the maximum scale $L$ of turbulent eddies around each cell and the local turbulent velocity $\delta\vecv(L)$ are obtained. 

\section{Unbound cloud}
\label{sc:blob}

\begin{figure*}
\centering
  \includegraphics[width=\linewidth]{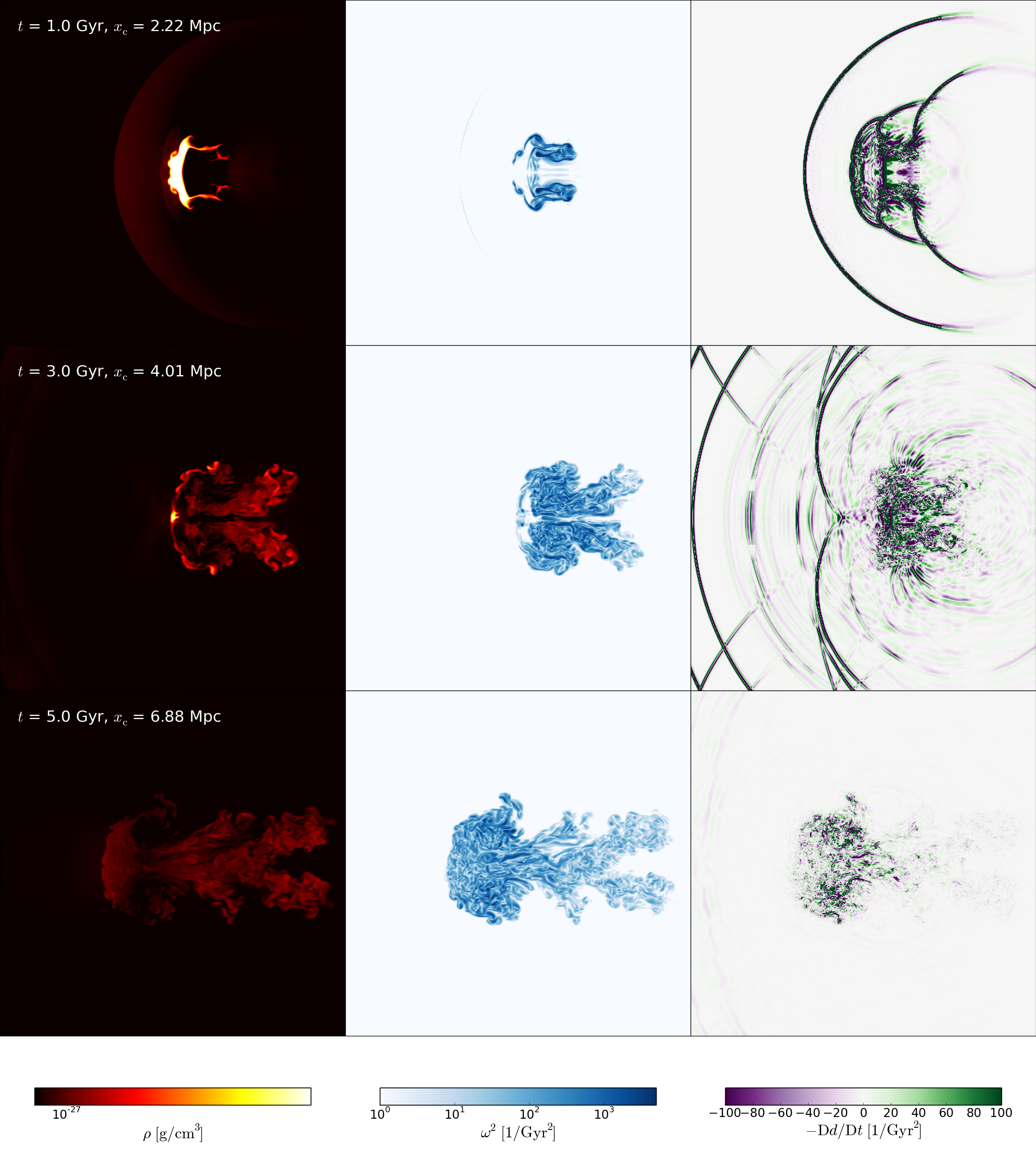}
\caption{
	Highest-resolution AMR simulation of the unbound cloud performed with the code {\sc Nyx}. Shown are 
	the mass density $\rho$ 
	(logarithmic scale), the squared vorticity modulus $\omega^2$, and the rate of compression
	$-\DD d/\dd t$ (see equation~\ref{eq:div}) in a $4\;\mathrm{Mpc}\times4\;\mathrm{Mpc}$ region
	of the $xy$-plane for 
	$t=1.0\;\mathrm{Gyr}$ (top), $3.0\;\mathrm{Gyr}$ (middle), and $5.0\;\mathrm{Gyr}$ (bottom).
	To follow the downstream drift of the cloud, the displayed region is centered at the coordinate
	$x_{\rm c}$ in $x$-direction.
	}
\label{fig:nyx_amr2x2}
\end{figure*}

The evolution of the unbound cloud in the AMR run with the highest resolution is illustrated in Figure~\ref{fig:nyx_amr2x2}. 
As reported by \citet{AgerMoore07}, the initially spherical cloud is rapidly compressed
into an oblate shape. Because of the supersonic wind, a bow shock forms in front of the cloud. 
Subsequently, a complex pattern of shocks and interfering pressure waves emerges, for which
the rate of compression is an excellent indicator.
After a few Gyr, the waves are mostly dissipated. One can also see the formation of vortices and the stripping of
gas from the cloud. This is a consequence of Kelvin-Helmholtz instabilities in the shear layers between the
background flow and the cloud interior and Rayleigh-Taylor instabilities due to the
acceleration of the denser cloud material by the low-density wind (see \citealt{AgerMoore07}
for a detailed discussion of these processes). As a result, a turbulent mushroom-like structures develops,
which is dragged by the wind in downstream direction. The density contrast gradually decreases because of the
mixing of the cloud material into the background medium.

\begin{figure*}
\centering
  \includegraphics[width=\linewidth]{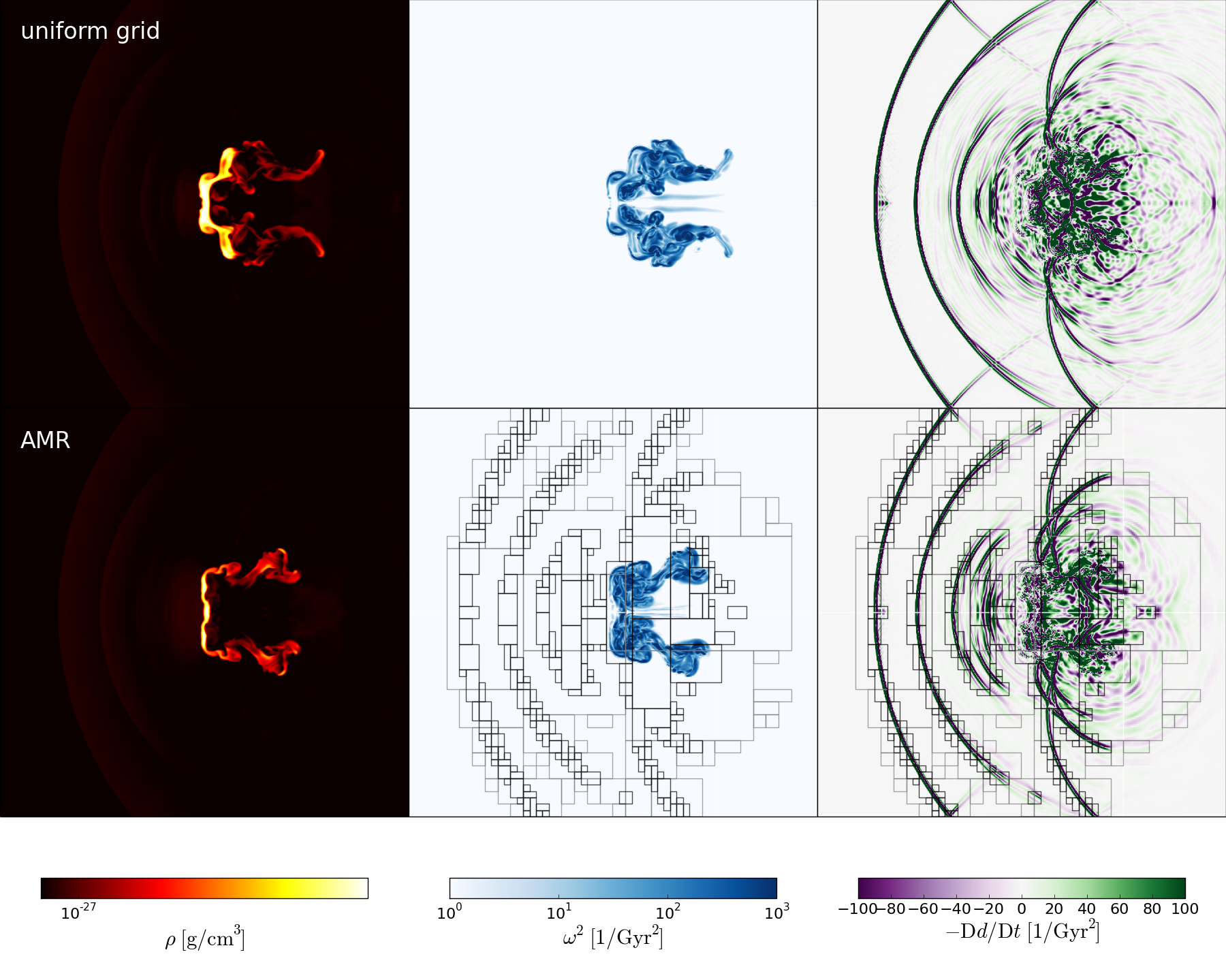}
\caption{
	Highest-resolution uniform-grid (top) and AMR (bottom) simulations of the unbound cloud performed with 
	the code {\sc Nyx}. Shown are 
	slices of the mass density $\rho$, the squared vorticity modulus $\omega^2$, and the rate of compression
	$-\DD d/\dd t$ for $t=2.0\;\mathrm{Gyr}$ (i.~e., a stage in between the top and middle rows in 
	Figure~\ref{fig:nyx_amr2x2}). 
	The boxes in the bottom panels indicate grid patches produced by AMR at level 1
	(gray boxes) and level 2 (black boxes). For example, from left to right the resolution increases upstream of the
	the first bow shock from level 0 to level 2 around the shock, then decreases to level 1 and increases
	again to level 2 around the second shock, etc.
	}
\label{fig:nyx_512_slices_2Gyr}
\end{figure*}

\begin{figure*}
\centering
  \includegraphics[width=\linewidth]{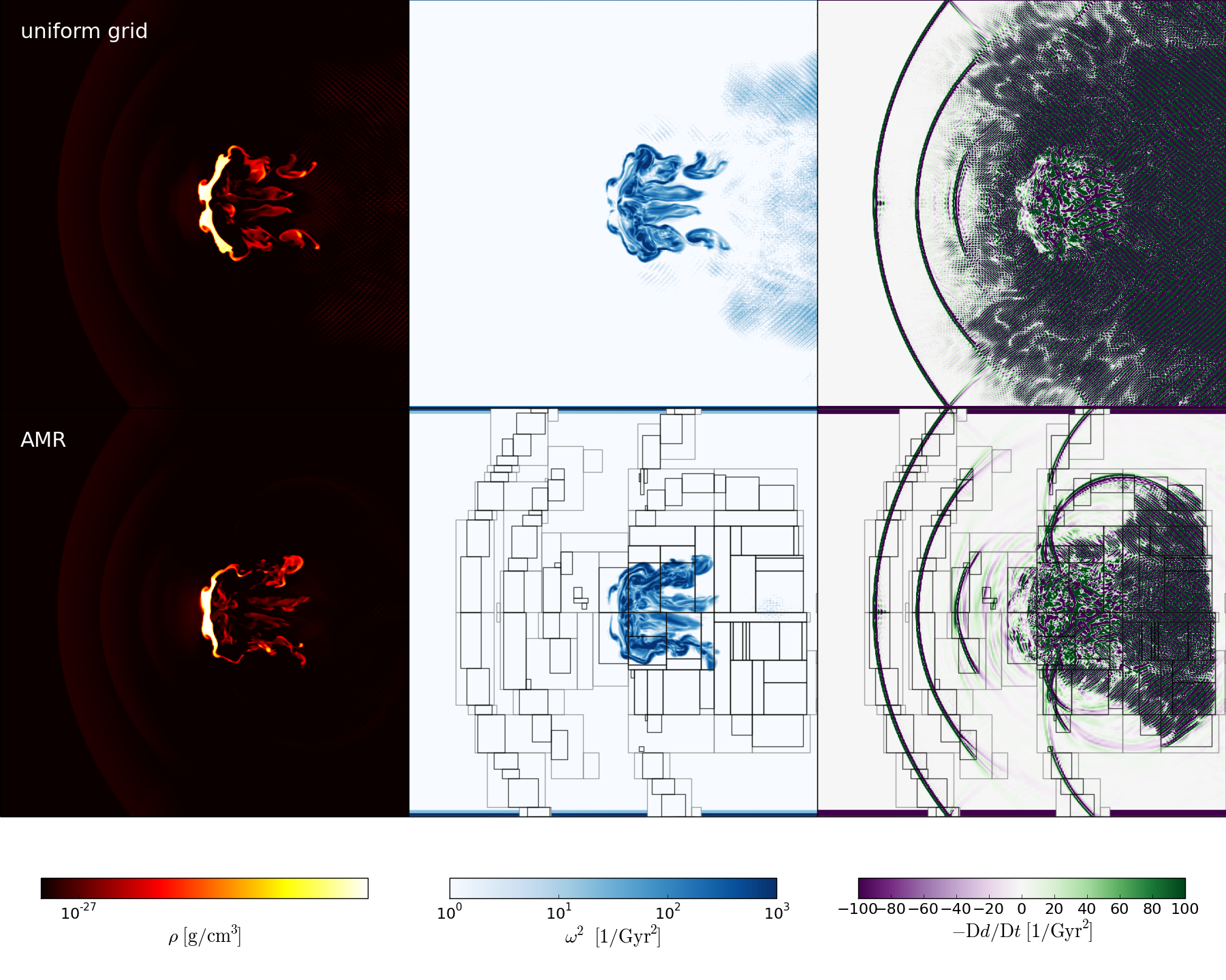}
\caption{
	The same as in Figure~\ref{fig:nyx_512_slices_2Gyr} for the corresponding simulations performed with
	{\sc Enzo}.}
\label{fig:enzo_512_slices_2Gyr}
\end{figure*}

\begin{figure*}
\centering
  \includegraphics[width=\linewidth]{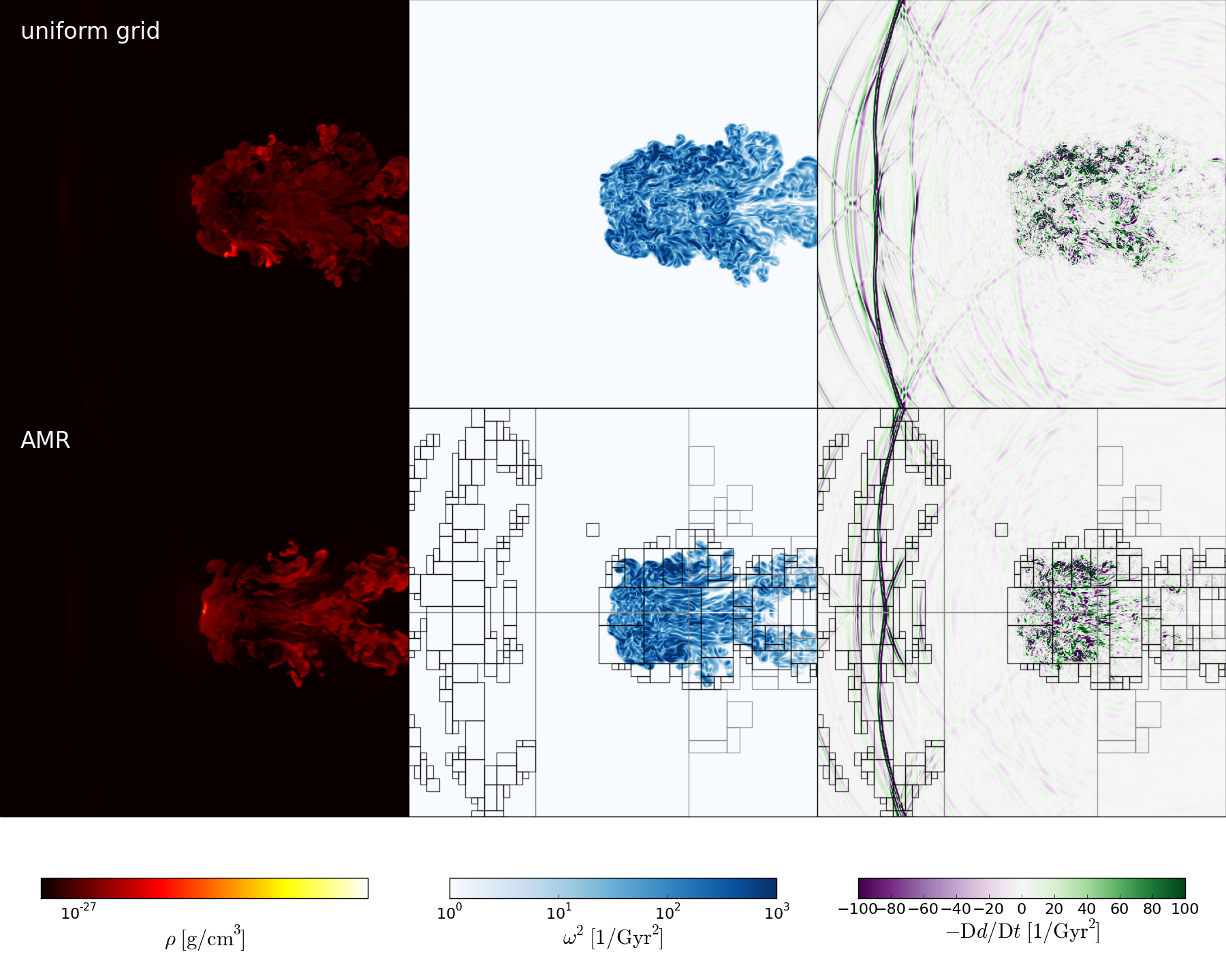}
\caption{
	The same as in Figure~\ref{fig:nyx_512_slices_2Gyr} for 
	the {\sc Nyx} simulation at} $t=4.0\;\mathrm{Gyr}$.
\label{fig:nyx_512_slices_4Gyr}
\end{figure*}

If a uniform grid with the same effective resolution as in the AMR run is used, the flow evolution is
qualitatively similar. However, the comparison between the uniform-grid and AMR runs at $t=2.0\;\mathrm{Gyr}$
in Figure~\ref{fig:nyx_512_slices_2Gyr} shows palpable differences in the cloud morphology (density slices) 
and the flow structure (vorticity slices). In particular, it appears that more gas is ablated from the 
frontal cap in the AMR run. 
Shocks, which appear as large arcs in the slices of the compression rate, are nearly identical in both runs. 
Slices from the corresponding {\sc Enzo} runs at the same instant of time are plotted in Figure~\ref{fig:enzo_512_slices_2Gyr}.
Compared to the {\sc Nyx} runs, the shape of the cloud is markedly different.
However, the deviation of the AMR run from the uniform-grid run 
appears to be less pronounced for {\sc Enzo}.\footnote{Although we implemented the same
	refinement criteria into {\sc Enzo} and {\sc Nyx}, one should bear in mind that the grid structure
	resulting from these criteria is not identical. Although both codes feature block-structured AMR, 
	there are fundamental differences in the grid generation and regridding algorithms
	and the associated control parameters (see \citealt{Enzo13}, \citealt{AlmBell13}, and the
	code documentations for details).} 
Deviations are, of course, inevitable because of the non-linear evolution of the flow instabilities, but the differences between 
the various runs are stronger than what one might expect. 
In the fully turbulent regime, after the cloud has
largely dissolved, the differences in the flow structure tend to be alleviated (see 
Figure~\ref{fig:nyx_512_slices_4Gyr} for $t=4.0\;\mathrm{Gyr}$). This is possibly
a consequence of the randomization caused by turbulence.

A major discrepancy between {\sc Nyx} and {\sc Enzo} that becomes apparent in the slices shown in Figures~\ref{fig:nyx_512_slices_2Gyr} 
and~\ref{fig:enzo_512_slices_2Gyr} is the interference pattern of waves in the surroundings of the cloud.
In both cases, the compression rate indicates pressure waves produced by the collision of the incoming gas
with the denser cloud. 
However, while the wavelength separating local maxima or minima of $-\DD d/\dd t$ is clearly larger for {\sc Nyx},
the wave amplitude is much smaller: $|\DD d/\dd t|\sim 10$ for {\sc Nyx}, but 
$|\DD d/\dd t|\gg 100$ for {\sc Enzo}. Apart from that, the wavelength apparently 
changes with the grid resolution in the case of Enzo (not shown here).
This suggests that the waves in the {\sc Enzo} runs might at least partially be of numerical origin.

\begin{figure*}
\centering
  \includegraphics[width=\linewidth]{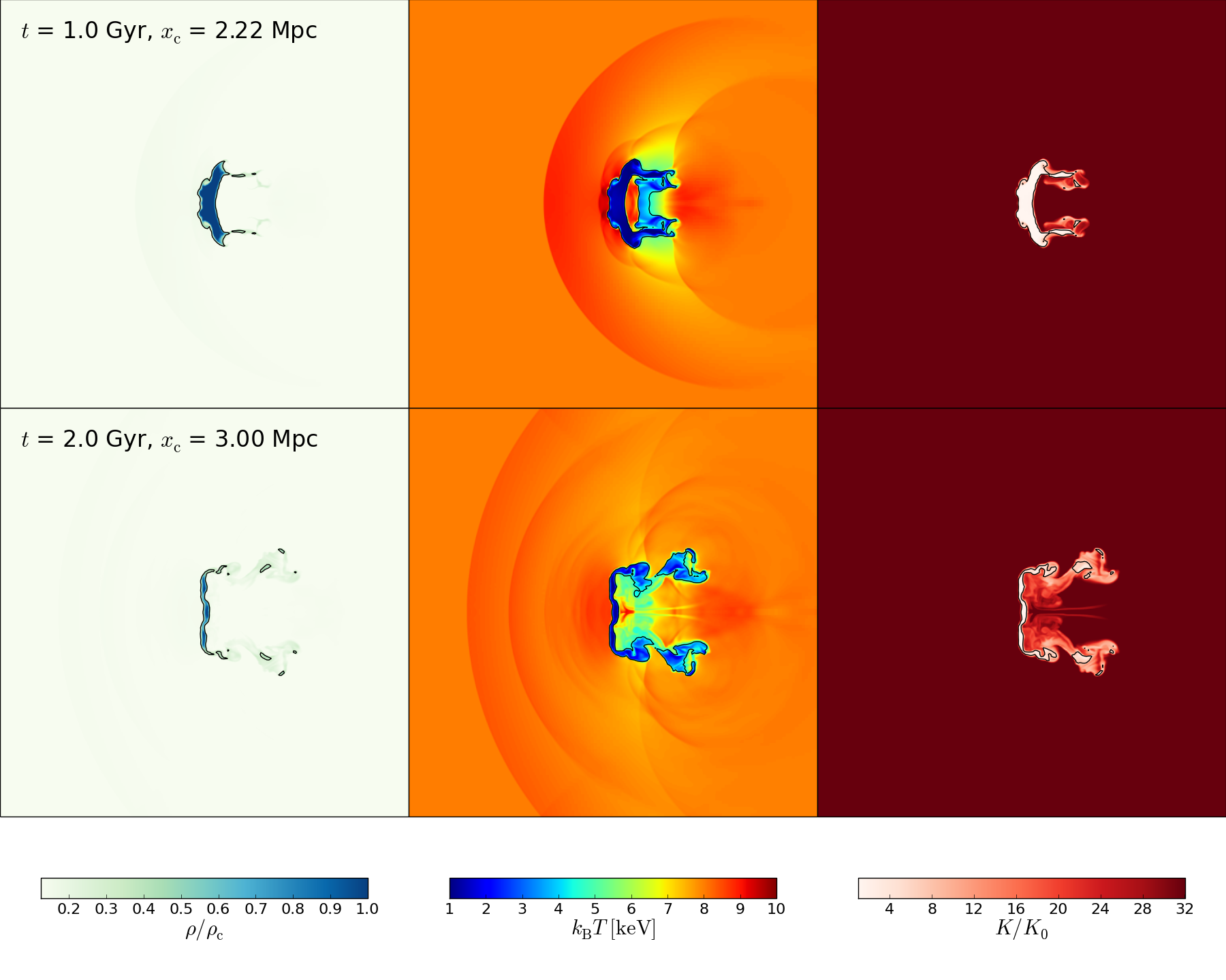}
\caption{
	Slices of the density, temperature, and entropy for the highest-resolution AMR simulation performed with {\sc Nyx}
	at two different instants. The density and entropy are normalized to the initial central density $\rho_{\rm c}$ 
	and entropy $K_{0}$, respectively, of the cloud. Isosurfaces for $\rho/\rho_{\rm c}=0.32$ (left),
	$T = 4.5\;\mathrm{keV}$ (middle), and $K/K_0=8.0$ (right) are indicated by black contour lines.
	The slices are centered at coordinates $x_{\rm c}$ in $x$-direction, as in Figures~\ref{fig:nyx_amr2x2} 
	and~\ref{fig:nyx_512_slices_4Gyr}.
	}
\label{fig:nyx_512_slices_entr}
\end{figure*}

\subsection{Mass stripping}
\label{sc:mass}

\begin{figure}
\centering
  \includegraphics[width=0.5\linewidth]{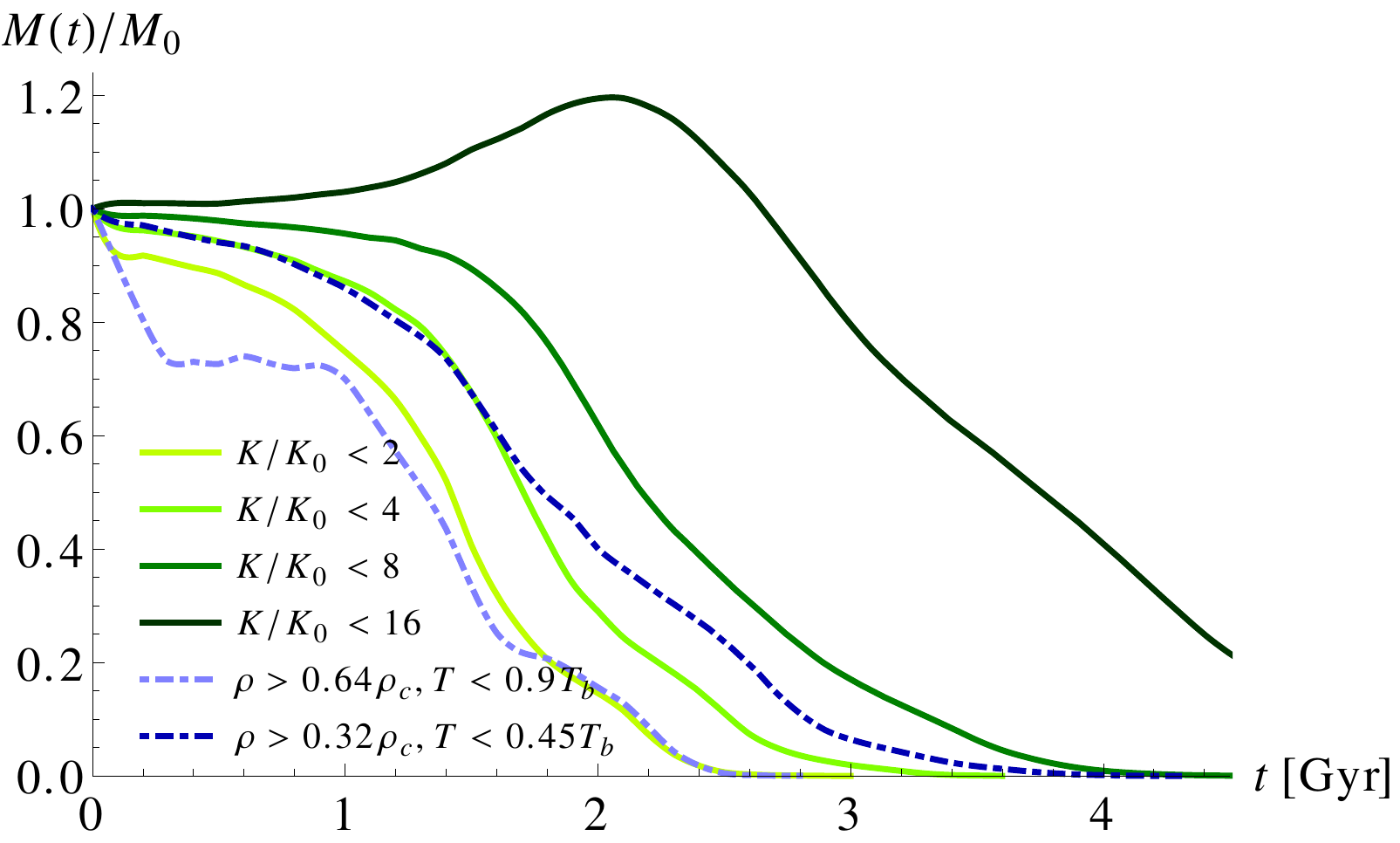}
\caption{Time evolution of the cloud mass for different choices of the entropy
	threshold (solid lines) for the {\sc Nyx} AMR simulation with one level of refinement.  
	The dot-dashed lines are obtained if the cloud is defined by a lower threshold for the
	density and an upper threshold for the temperature.}
\label{fig:mass_compr}
\end{figure}

An important question is whether robust statistics, which are not too sensitive to the
numerical realization, can be inferred from the simulations. To characterize the mass stripping from the cloud,
\citet{AgerMoore07} and \citet{IapiAda08} use the total mass of gas that has temperatures
below and densities above given thresholds as statistical measure.
The density and temperature thresholds chosen by \citet{AgerMoore07} are $0.64\rho_{\rm c}$ and
$0.9T_{\rm b}$, while \citet{IapiAda08} set a lower threshold of $0.32\rho_{\rm c}$ for the density.
The regions with $\rho/\rho_{\rm c}\ge 0.32$ at $t=1.0$ and $2.0\;\mathrm{Gyr}$ are shown for the
{\sc Nyx} AMR simulation with two levels of refinement in Figure~\ref{fig:nyx_512_slices_entr}. The temperature
slices in the middle make clear that the threshold $0.9k_{\rm B}T_{\rm b}= 7.2\;\mathrm{keV}$
encloses a lot of low-density material from the ambient medium, which is heated by adiabatic compression. 
For this reason, we consider a lower maximal temperature of $T_{\rm c}+0.5(T_{\rm b}-T_{\rm c})\approx 0.56T_{\rm b}$, 
corresponding to $4.5\;\mathrm{keV}$, to constrain the cloud interior. 
The resulting time evolution of the total mass $M(t)$
of the gas with $\rho/\rho_{\rm c}\ge 0.32$ and $k_{\rm B}T\le 4.5\;\mathrm{keV}$ is plotted as dot-dashed line
in Figure~\ref{fig:mass_compr}.\footnote{Since the data required to calculate $M(t)$ are very large
	for the highest-resolution case, we used the lower-resolution AMR run for the comparison in
	Figure~\ref{fig:mass_compr}.} 
At $t=1\;\mathrm{Gyr}$, the cloud is compressed and deformed (see Figure~\ref{fig:nyx_512_slices_entr}), 
but still retains a large fraction of the initial mass. During a phase of rapid mass loss, the cloud is fragmented
and about half of its initial mass is mixed into the background medium. 
Eventually, after about $3\;\mathrm{Gyr}$, the cloud has lost most of its mass.
Compared to the thresholds $0.64\rho_{\rm c}$ and $7.2\;\mathrm{keV}$ as in \citet{AgerMoore07}, 
$M(t)$ initially decreases more gradually and the lifetime of the cloud is longer.

In contrast to the temperature, the entropy is not increased by adiabatic compression (with the exception of
shocks). However, the entropy of the gas in the cloud is raised as it is mixed into the background
medium by turbulence. This suggests that the cloud interior can be defined by an entropy threshold. Indeed,
the low-entropy material of the cloud stands out against the background medium in the slices of the
entropy shown in Figure~\ref{fig:nyx_512_slices_entr}. Thus, we define the cloud interior by the criterion
\begin{equation}
	\label{eq:entropy}
	K=\frac{T}{\rho^{2/3}}\le K_{\rm max}=f K_0,
\end{equation}
where $K_0=T_{\rm c}/\rho_{\rm c}^{2/3}$ is the initial entropy in the centre of the cloud and $f>1$ a dimensionless factor. 
The cloud masses defined by the above criterion with $f=2$, $4$, $8$, and $16$ are plotted as functions of
time in Figure~\ref{fig:mass_compr}. The cloud mass defined by $\rho/\rho_{\rm c}\ge 0.32$ and $k_{\rm B}T\le 4.5\;\mathrm{keV}$
falls in between the cases $f=4$ and $f=8$. For $f=2$, there is a sudden drop of the cloud mass just after $t=0$, 
while $f=16$ does not sufficiently constrain the cloud interior to obtain a monotonously decreasing mass. 
We choose $f=8$ as fiducial value for our analysis because the mass is smoothly ramped down over a relatively long
period of $4\;\mathrm{Gyr}$ (the corresponding entropy contours are overplotted in Figure~\ref{fig:nyx_512_slices_entr}).

\begin{figure*}
\centering
  \includegraphics[width=0.48\linewidth]{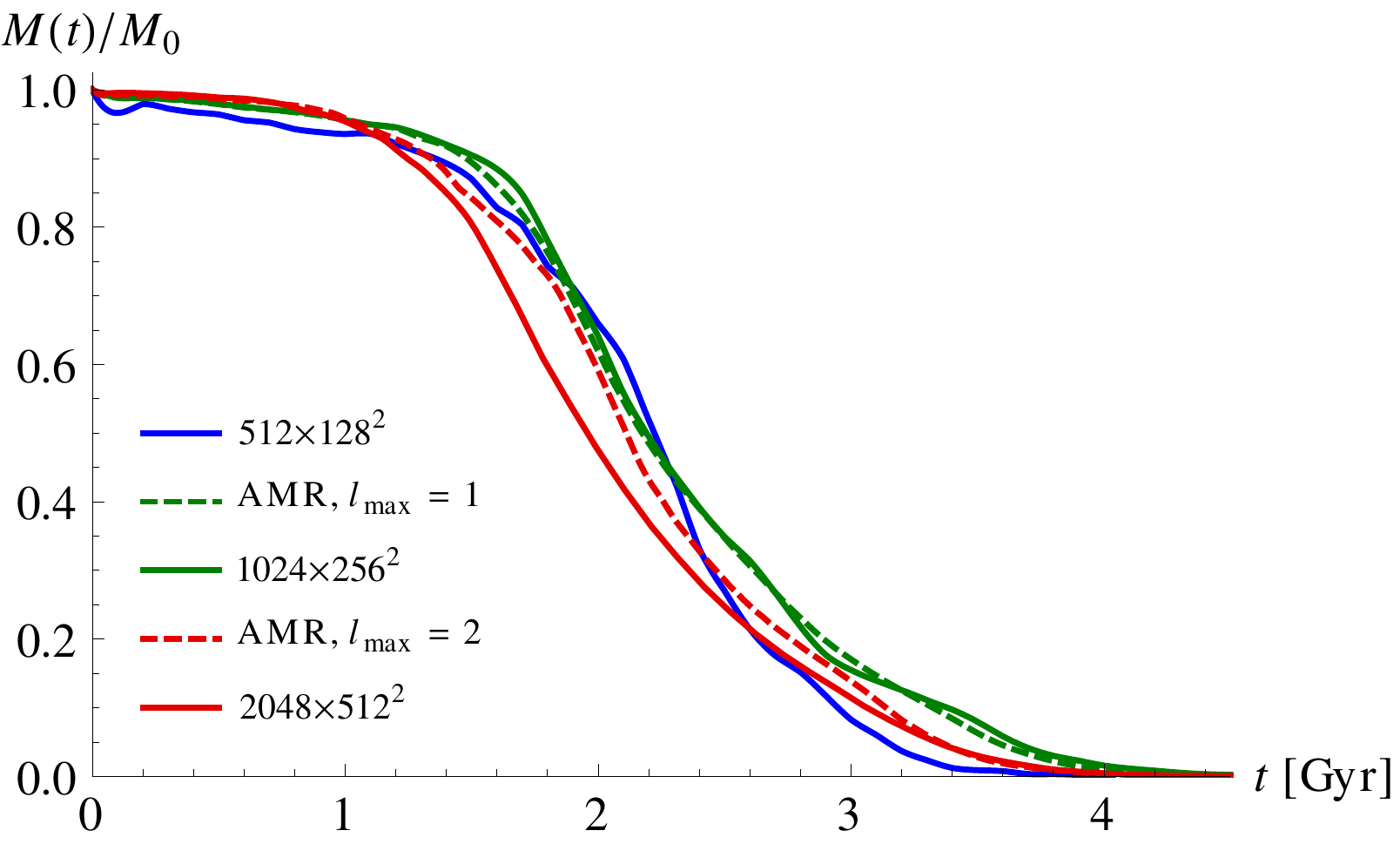}
  \includegraphics[width=0.48\linewidth]{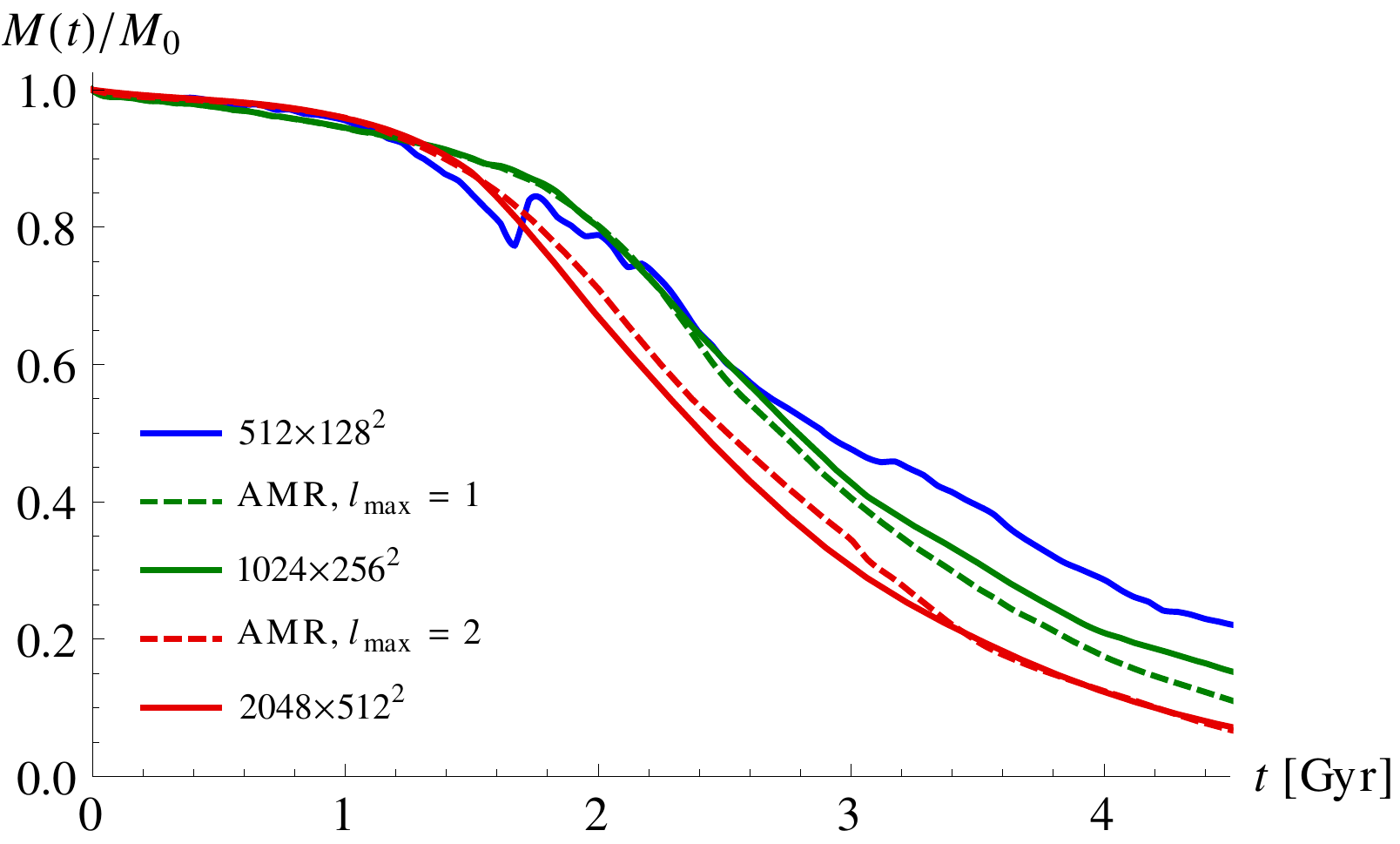}
\caption{Time evolution of the cloud mass in AMR simulations (dashed lines) with $n_{\rm max}$ levels of refinement
      and in uniform-grid simulations (solid lines) with different resolutions for {\sc Nyx} (left) and {\sc Enzo} (right).
      Simulations with the same effective resolutions are shown in the same colours (online version).
      For all curves, the cloud interior is defined by $K/K_0<8$.}
\label{fig:mass}
\end{figure*}

The mass statistics for the simulations listed in the upper part of Table~\ref{tb:runs} are plotted in Figure~\ref{fig:mass}.
One can see in the left plot that the mass stripping in the {\sc Nyx} simulations is relatively robust for different resolutions
as well as uniform-grid vs.\ AMR simulations, although there is a relatively large deviation of the run with a uniform grid 
at maximal resolution ($\Delta_{\rm min}=7.8\;\mathrm{kpc}$),
particularly in the early mass-loss phase between $1$ and $2\;\mathrm{Gyr}$. This might be a consequence of the differences 
in the cloud morphology that can be seen in Figure~\ref{fig:nyx_512_slices_2Gyr}. The less pronounced differences in the corresponding
 {\sc Enzo} simulations (see Figure~\ref{fig:enzo_512_slices_2Gyr}) are reflected by a better
agreement of $M(t)$ between the runs with the same effective resolution (right plot in Figure~\ref{fig:mass}). 
However, not only is the dependence on effective resolution stronger in the case of {\sc Enzo}, but there are striking
systematic differences: The mass stripping progresses clearly faster if {\sc Nyx} is used. 
This behavior is further investigated for the subcluster in Section~\ref{sc:msc}. 

\begin{figure*}
\begin{center}
\includegraphics[width=0.99\textwidth]{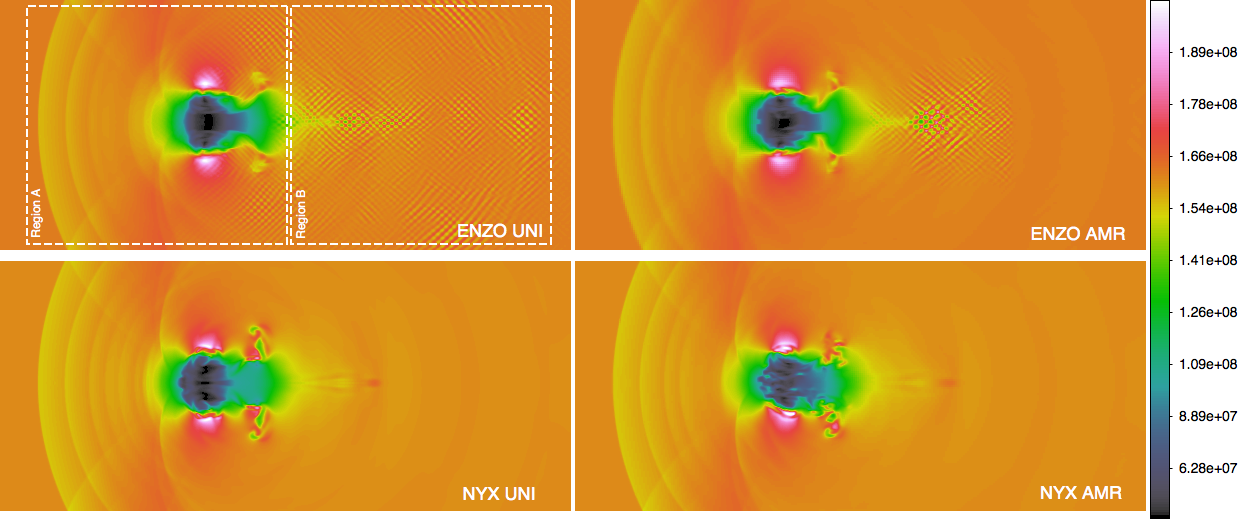}
\caption{Maps of total velocity modulus (in units of cm/s) for a slice crossing the middle plane of the unbound cloud runs at time 
$t=3\;\mathrm{Gyr}$. The top panels show the maps for {\sc Enzo} with a uniform grid of $1024 \times 256^2$ cells (left) and with AMR,
using 1 level of refinement (right). The bottom panels show the corresponding results for {\sc Nyx}. The additional dashed selections display
the subregions used for our analysis of turbulent velocity statistics.}
\label{fig:velocity}
\end{center}
\end{figure*}

\begin{figure*}
\begin{center}
\includegraphics[width=0.99\textwidth]{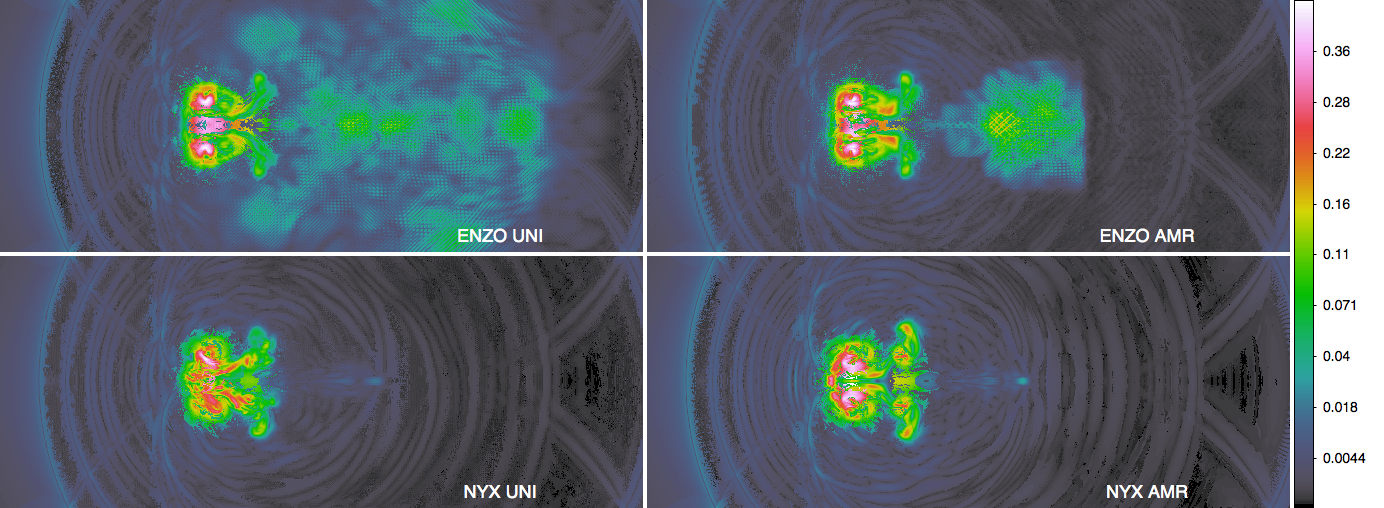}
\caption{Maps of turbulent velocity modulus relative to the upstream wind velocity $v_{\rm b}$
	for the same runs and selections as in Fig.~\ref{fig:velocity}.}
\label{fig:filter1}
\end{center}
\end{figure*}

\subsection{Turbulent velocity}

To obtain a better understanding of the differences between the codes, we analyze turbulent flow statistics.
Figure \ref{fig:velocity} shows the total velocity map for a slice through the middle plane of the unbound cloud runs 
(we consider here the intermediate resolution of $1024 \times 256^2$ and the corresponding AMR run with 1 level of refinement) 
at the epoch of $3\;\mathrm{Gyr}$. Apart from the shock waves in the upstream region, the velocity field in the uniform-grid 
{\sc Enzo} run shows a prominent pattern of small-scale perturbations, which fills a much larger volume than 
in the corresponding AMR case. This pattern also can be seen in the compression rate (not plotted here).
The fluctuations are better highlighted by the filtering of velocity data with the algorithm presented 
in Section \ref{sc:methods}. For the same instant of time, Figure~\ref{fig:filter1} presents slices showing
the modulus of the turbulent field $\delta\vecv(L)$ resulting from the filtering algorithm. As expected, 
the filtering procedure reveals a "noisy" pattern of interfering small-scale waves in {\sc Enzo}, just downstream of the blob 
in the AMR run and even at the sides of the blob in the run with a uniform grid. 
These small-scale waves are totally absent in {\sc Nyx} runs, 
where only a coherent pattern of sound waves originating from the cloud can be seen. 
Another interesting difference between the two codes is that
the turbulent velocity field in the cloud interior and wake appears to be more irregular in the {\sc Nyx} run. 
The higher symmetry for {\sc Enzo} is possibly a consequence of the propensity of Kelvin-Helmholtz rolls to get aligned
with the grid if directional splitting is used. Since the unsplit solver implemented in {\sc Nyx} reduces the influence 
of the grid directions, it is conceivable that the turbulent flow randomizes faster. This could also explain the
faster mass stripping in the {\sc Nyx} runs (see Section~\ref{sc:mass}).

\begin{figure}
\begin{center}
\includegraphics[width=0.5\textwidth]{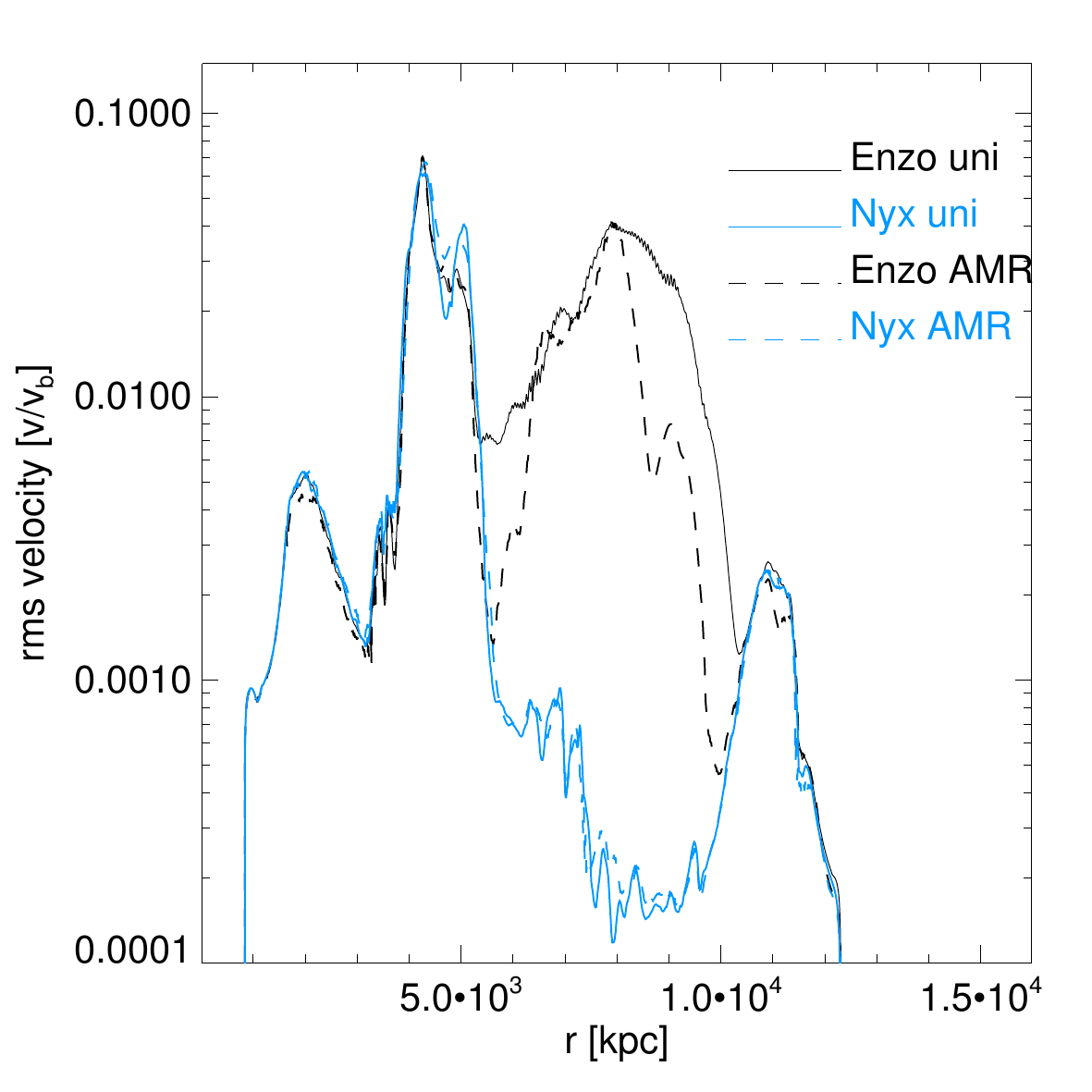}
\caption{Profiles of the root mean square velocities within slices perpendicular to the wind direction for the same datasets as
	in Figure~\ref{fig:filter1} ($t=3\;\mathrm{Gyr}$). 
	The velocities are normalised to the upstream wind velocity, $v_{\rm b}$, in both cases.}
\label{fig:filter2}
\end{center}
\end{figure}

\begin{figure}
\begin{center}
\includegraphics[width=0.5\textwidth]{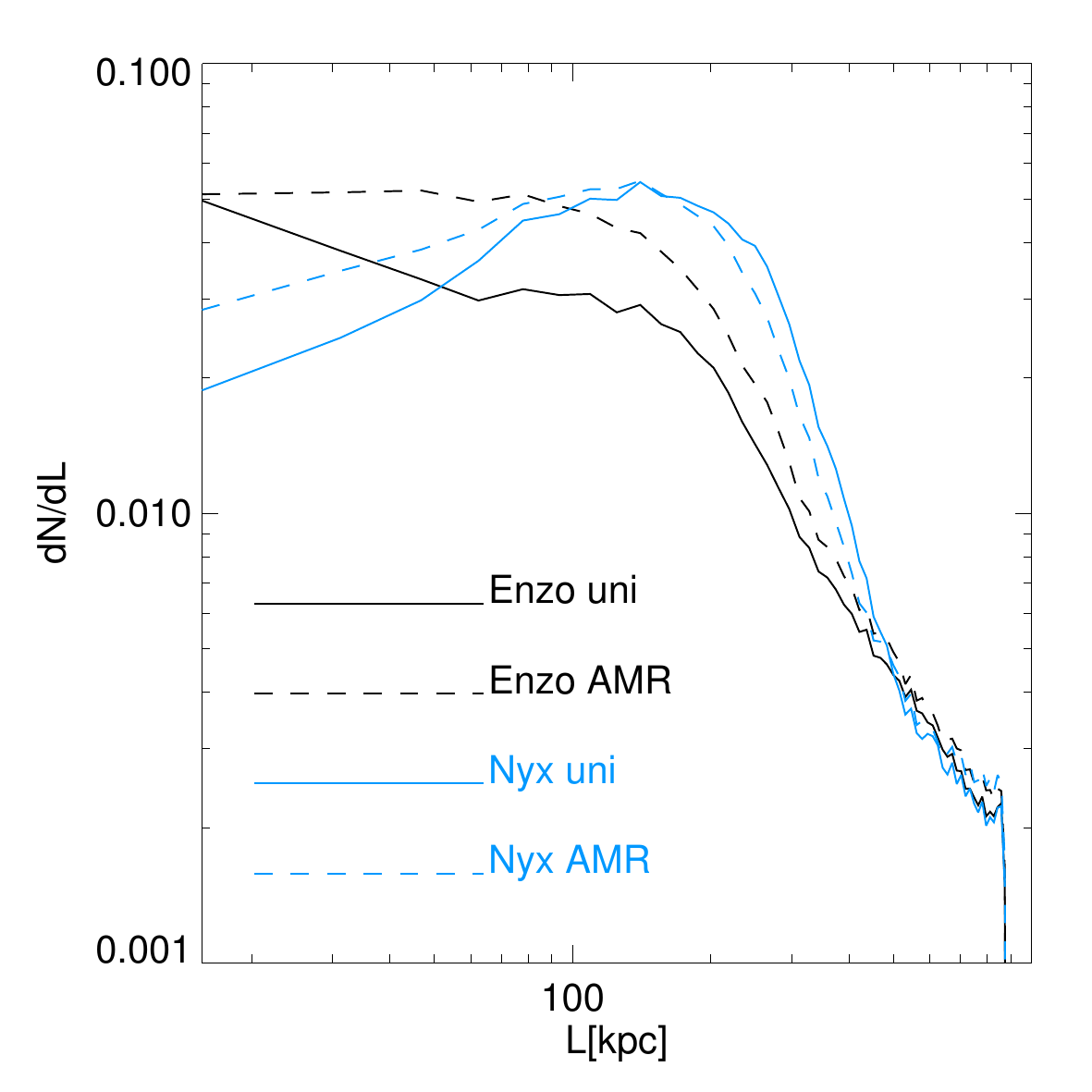}
\caption{Volume fractions of correlation scales in the flow (as reconstructed by the filtering algorithm explained in Section \ref{sc:methods}) for the datasets shown in Figure~\ref{fig:filter1} ($t=3\;\mathrm{Gyr}$).}
\label{fig:filter3}
\end{center}
\end{figure}

The turbulent velocity profiles plotted in Fig.~\ref{fig:filter2} are obtained by calculating the root mean square (rms) velocity for 
slices perpendicular to the wind direction. 
While the profiles agree within a few percent for most of the volume, the {\sc Enzo} data show a striking excess of velocity dispersion 
downstream of the cloud. The corresponding region is marked as region B in Fig.~\ref{fig:velocity}.
Figure~\ref{fig:filter3} shows the volume distribution of correlation scales $L$ in the flow, as given by 
the filtering algorithm of Section~\ref{sc:methods}. In comparison to {\sc Nyx}, 
the {\sc Enzo} run appears to have an excess of small-scale structure, which can be associated with the "noisy" pattern in Fig.~\ref{fig:filter1}. 
For scales larger than about 25 cells, the two codes produce very similar distributions, confirming that the largest scales in the flow are reconstructed in a very similar way. 

\begin{figure*}
\begin{center}
\includegraphics[width=0.49\textwidth]{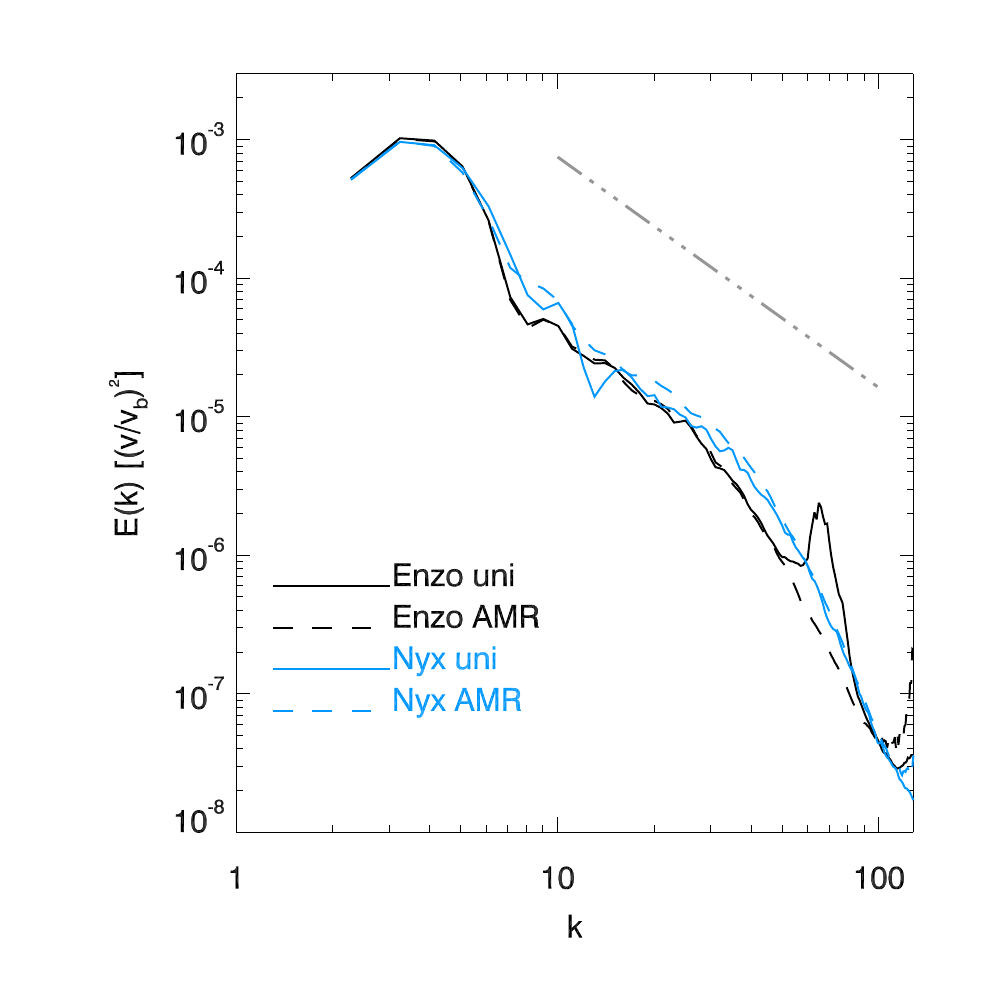}
\includegraphics[width=0.49\textwidth]{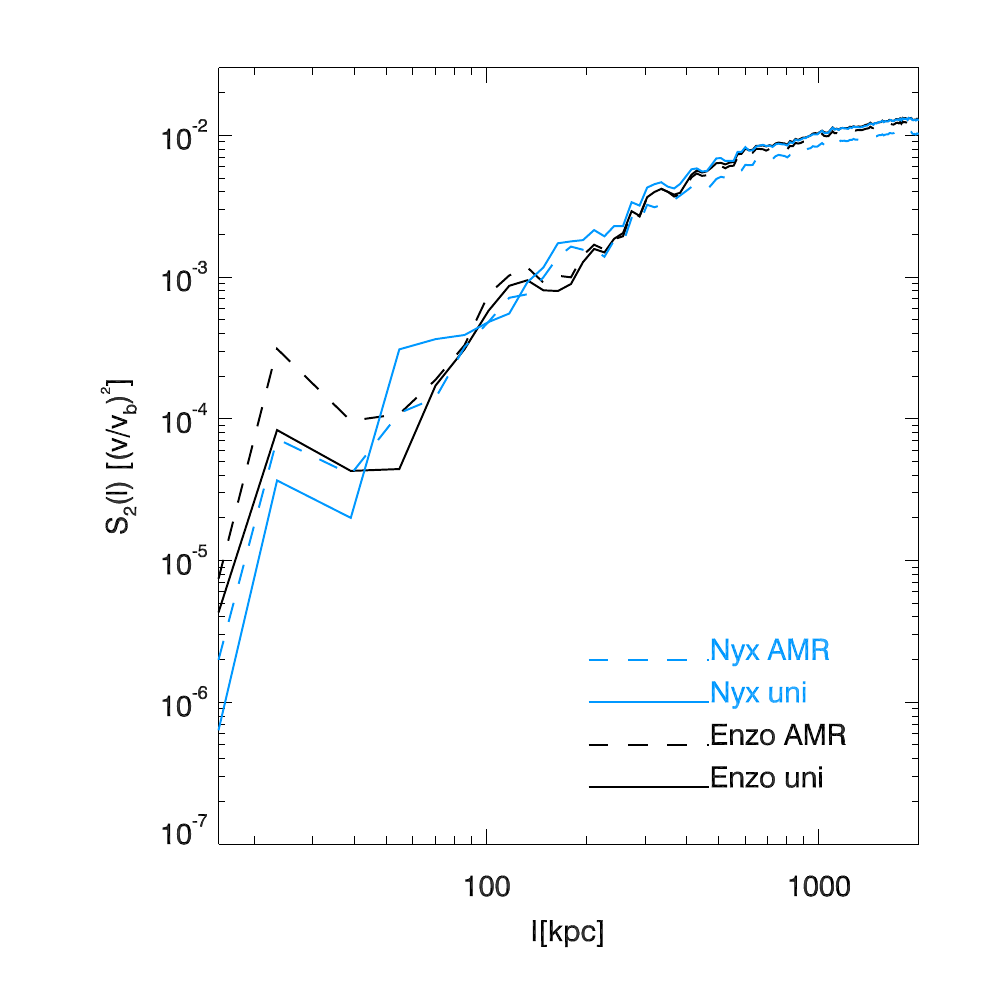}
\includegraphics[width=0.49\textwidth]{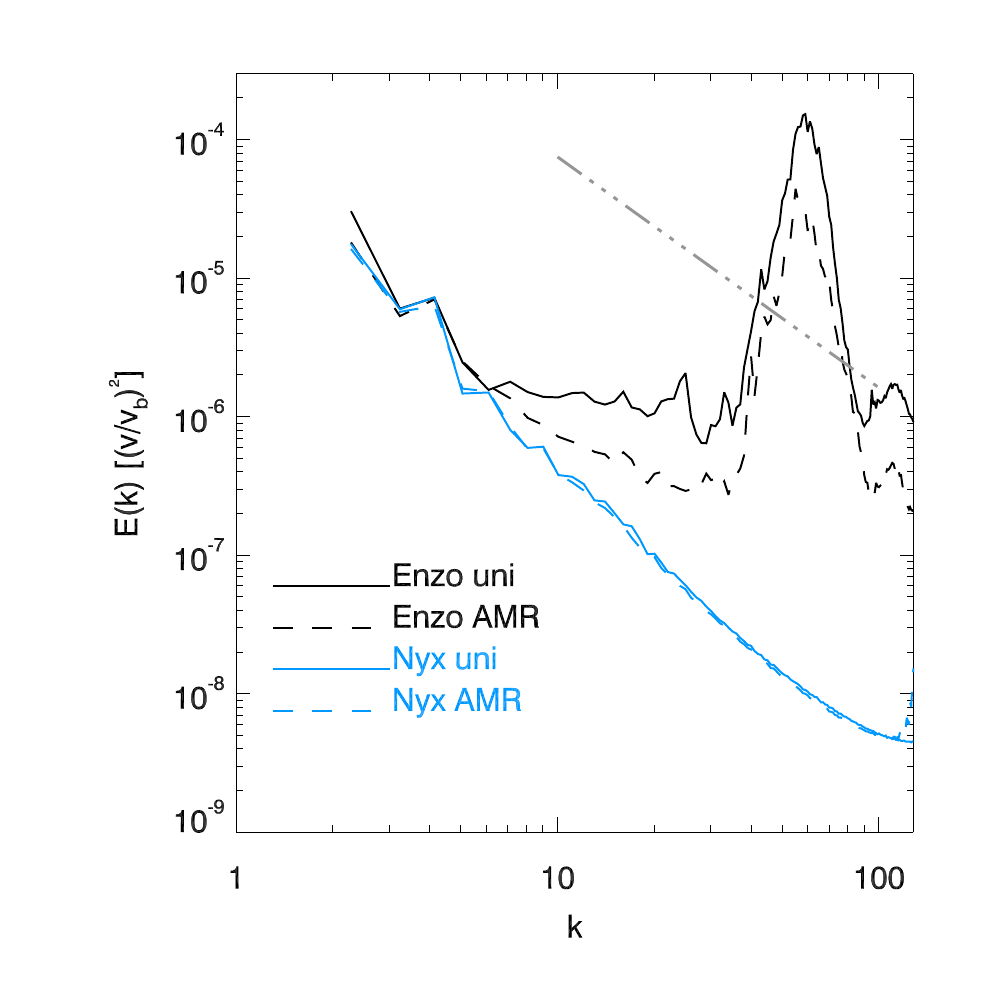}
\includegraphics[width=0.49\textwidth]{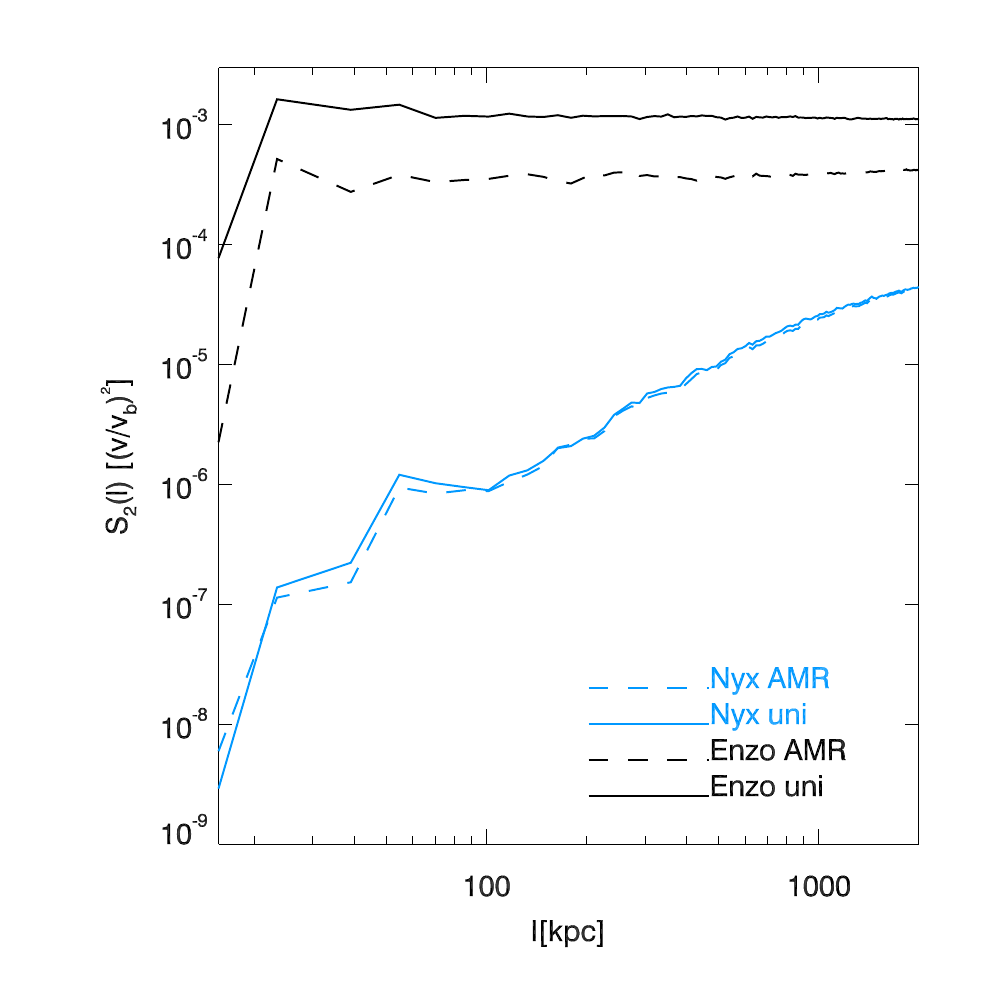}
\caption{Power spectra of the velocity field (left panels) and second order structure functions (right panels) for two subvolumes of the
computational domain at time $t=3.0\;\mathrm{Gyr}$ for the various {\sc Enzo} and {\sc Nyx} runs with a resolution of $15.6\;\mathrm{kpc}$ 
(see Table~\ref{tb:runs}). Region A (top panels) is cloud-centered, while region B (bottom panels) contains the noisy pattern observed downstream in the Enzo run. The dot-dashed grey line shows the Kolmogorov slope ($k^{-5/3}$).}
\label{fig:filter4}
\end{center}
\end{figure*}

We further investigated the properties of the velocity field by computing 3D power spectra and second-order structure functions of the total (unfiltered) velocity field, defined as $S_2(l) = \langle|\delta\vecv(l)|^2 \rangle$, where $\delta\vecv(l)$ is the velocity difference between two points separated by a distance $l$. For the spectra and structure functions, we selected two cubic subregions with $256^3$ cells 
in the simulation domain, as shown in Fig.~\ref{fig:velocity}. 
Region A contains the cloud and most of the turbulent wake, whereas the noisy pattern detected in the {\sc Enzo} runs is encompassed by region B. 
The structure functions were computed for $10^8$ randomly chosen pairs of cells within each region, while the power spectra were computed with a standard FFT approach, assuming periodic boundary conditions. The results are shown in Fig.~\ref{fig:filter4}. 
For region A, the power spectra and structure functions obtained with the different codes are in good agreement for low to
intermediate wavenumbers and length scales larger than a few $\Delta_{\rm min}$, respectively, regardless of using AMR or a uniform grid.
However, there is a pronounced spike around $k \sim 70$, where $k=1$ corresponds to the size of the whole region. 
This indicates the onset of small-scale velocity fluctuations around the cloud in the uniform-grid {\sc Enzo} run 
(see left top plot in Fig.~\ref{fig:velocity}). 
The slope of the spectra is roughly consistent with a Kolmogorov spectrum for intermediate wave numbers. However, the 
structure functions show no indication of a power law at all. Since the flow in subregion A is a mixture of turbulence and the surrounding wind,
a clean power law cannot be expected. Apart from that, the numerical resolution is too low to resolve inertial-range scaling
in the turbulent wake. 

Even more striking differences between {\sc Enzo} and {\sc Nyx} are measured in region B
(bottom plots in Fig.~\ref{fig:filter4}), where the noisy pattern of small-scale waves observed in Fig.~\ref{fig:velocity} 
produces a very prominent peak for wavenumbers $k \sim 30$ to $100$ (i.~e., scales of $2$ to $5$ grid cells). 
Since there are only pressure waves, but hardly any turbulence in this region, the spectrum should fall off steeply with
the wave number, which is indeed seen for {\sc Nyx}. The differences between the second-order structure functions
in region B (lower right plot in Fig.~\ref{fig:filter4}) are even more pronounced. Compared to {\sc Nyx}, 
the structure functions calculated from the {\sc Enzo} data are almost flat and much larger in magnitude. 
However, the peak-like features that can be seen in the power spectra are not present. Possibly, the rapid fluctuations associated with the small-scale waves tend to be correlated over a few cells such that their contribution to $\delta\vecv(l)$ is suppressed. Moreover, the structure functions are very coarse for low $l$ compared to the large number of narrow wavenumber bins for high $k$. This might also obscure the peaks. There is no
doubt, however, that the two codes produce substantially different levels of non-turbulent, noisy fluctuations, particularly in region B.

\begin{figure*}
\centering
  \includegraphics[width=\linewidth]{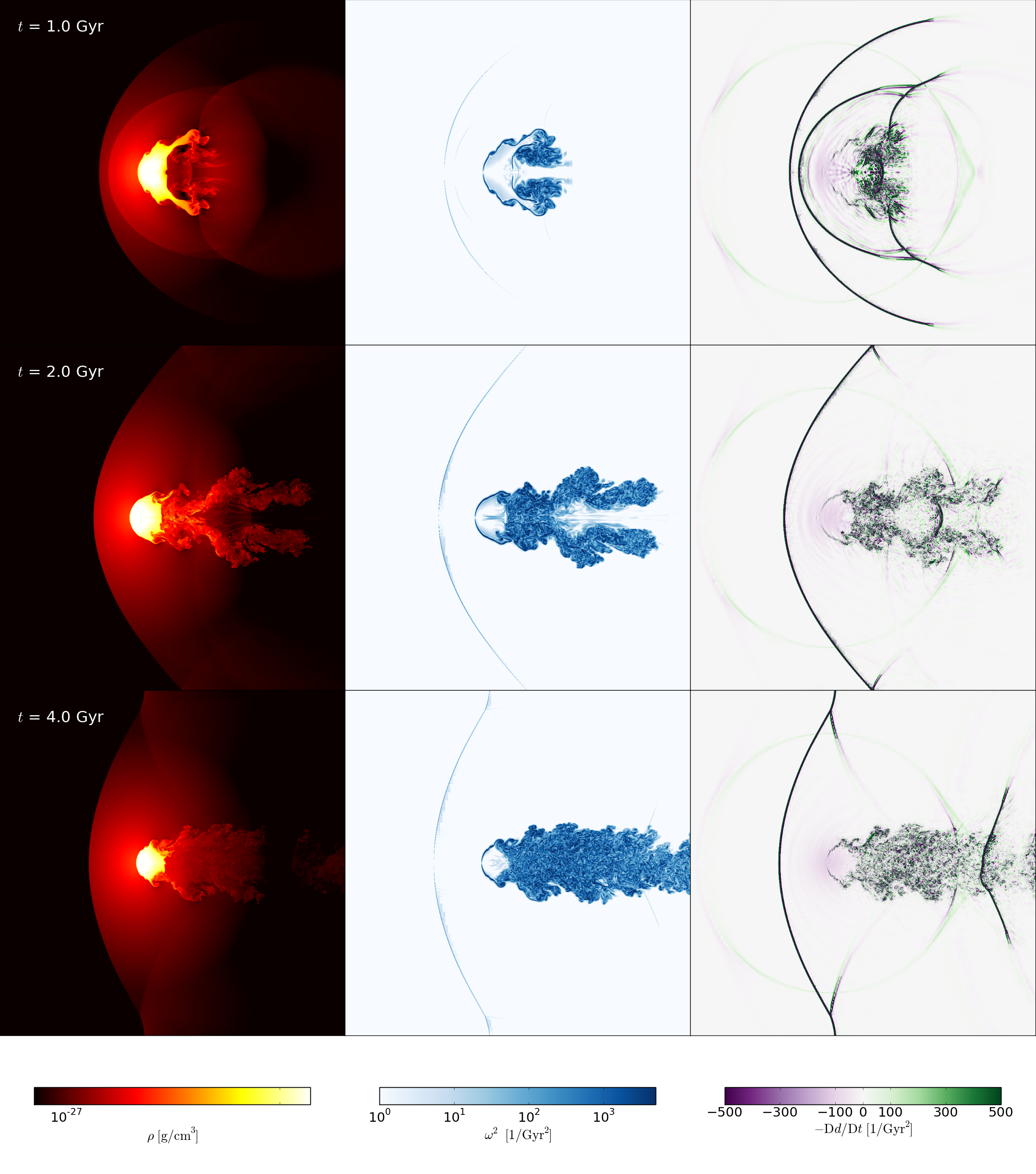}
\caption{
	Highest-resolution AMR simulation of the subcluster performed with the code {\sc Nyx}. Shown are 
	slices of the mass density $\rho$, the squared vorticity modulus $\omega^2$, and the rate of compression
	$-\DD d/\dd t$ for $t=1.0\,\mathrm{Gyr}$ (top), $2.0\,\mathrm{Gyr}$ (middle), and $4.0\,\mathrm{Gyr}$ (bottom).
	}
\label{fig:grav_nyx_amr3x2}
\end{figure*}

\section{Subcluster}
\label{sc:msc}

The evolution of the subcluster, i.~e., the bound cloud in a static gravitational potential, is illustrated in Fig.~\ref{fig:grav_nyx_amr3x2} 
for the {\sc Nyx} AMR simulation with three levels of refinement. As in the case of the unbound cloud, one can see that the
initially spherical subcluster is deformed by the ram-pressure of the wind after $1\;\mathrm{Gyr}$ (top panels) and mass is stripped from the
cloud by vortex shedding. However, the subsequent evolution is very different, as the subcluster is anchored by gravity. 
The gravitational potential produces a negative contribution to the rate of compression (see equation~\ref{eq:div}), which is 
visible near the center in the right panels. Also the cutoff of the potential at radius $r_{\rm max}$ can be discerned as a faint circle. 
Much stronger compression is caused by shocks and compressible turbulent fluctuations. At $t=2\;\mathrm{Gyr}$ (middle panels), 
there is a nearly stationary bow shock in front of the cloud and a turbulent wake begins to form. The final state at 
$t=4\;\mathrm{Gyr}$,  when the simulation is terminated, is shown in the bottom panels. At this stage, the flow in downstream direction
behind the cloud is fully turbulent. 

\begin{figure*}
\centering
  \includegraphics[width=0.48\linewidth]{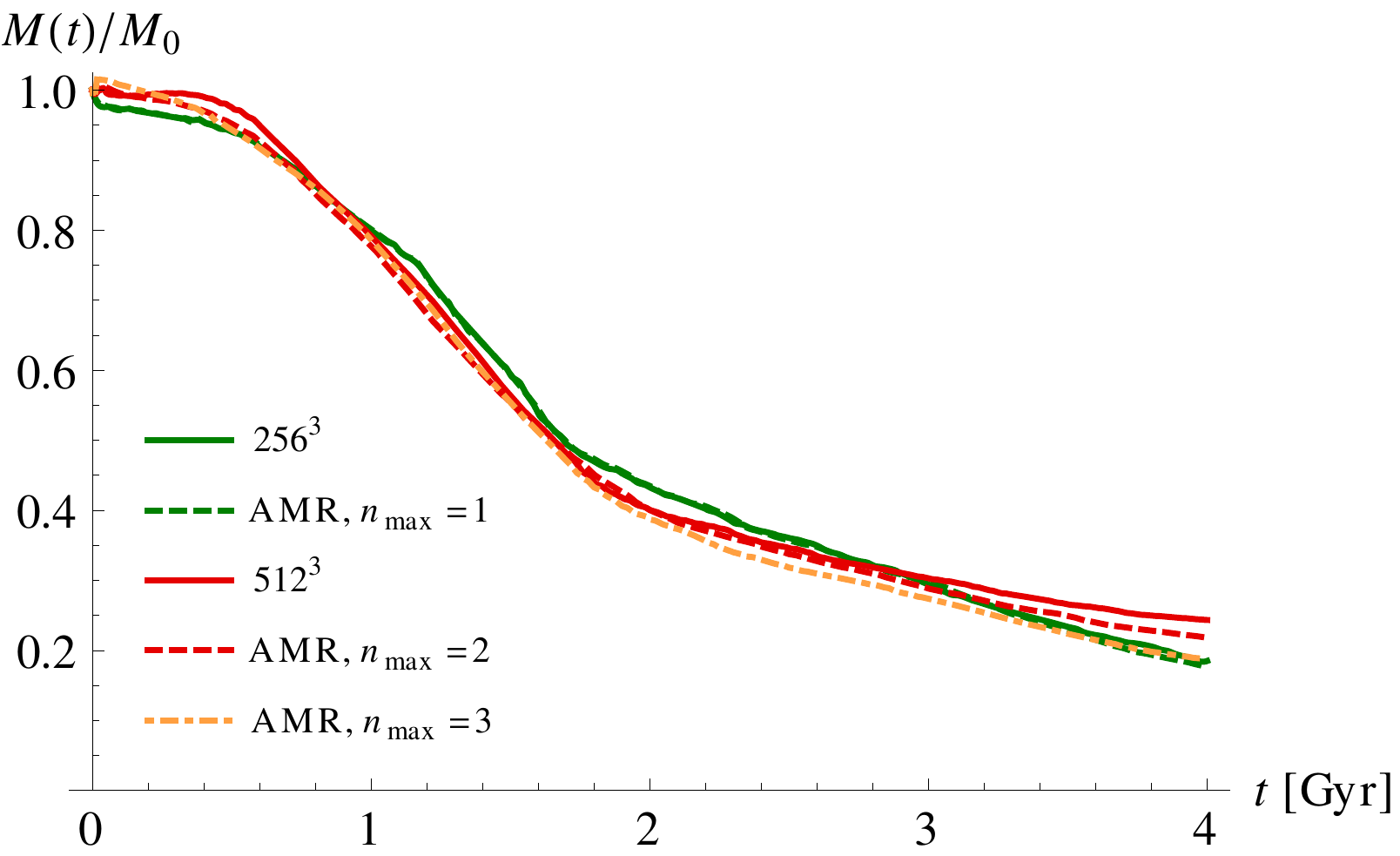}
  \includegraphics[width=0.48\linewidth]{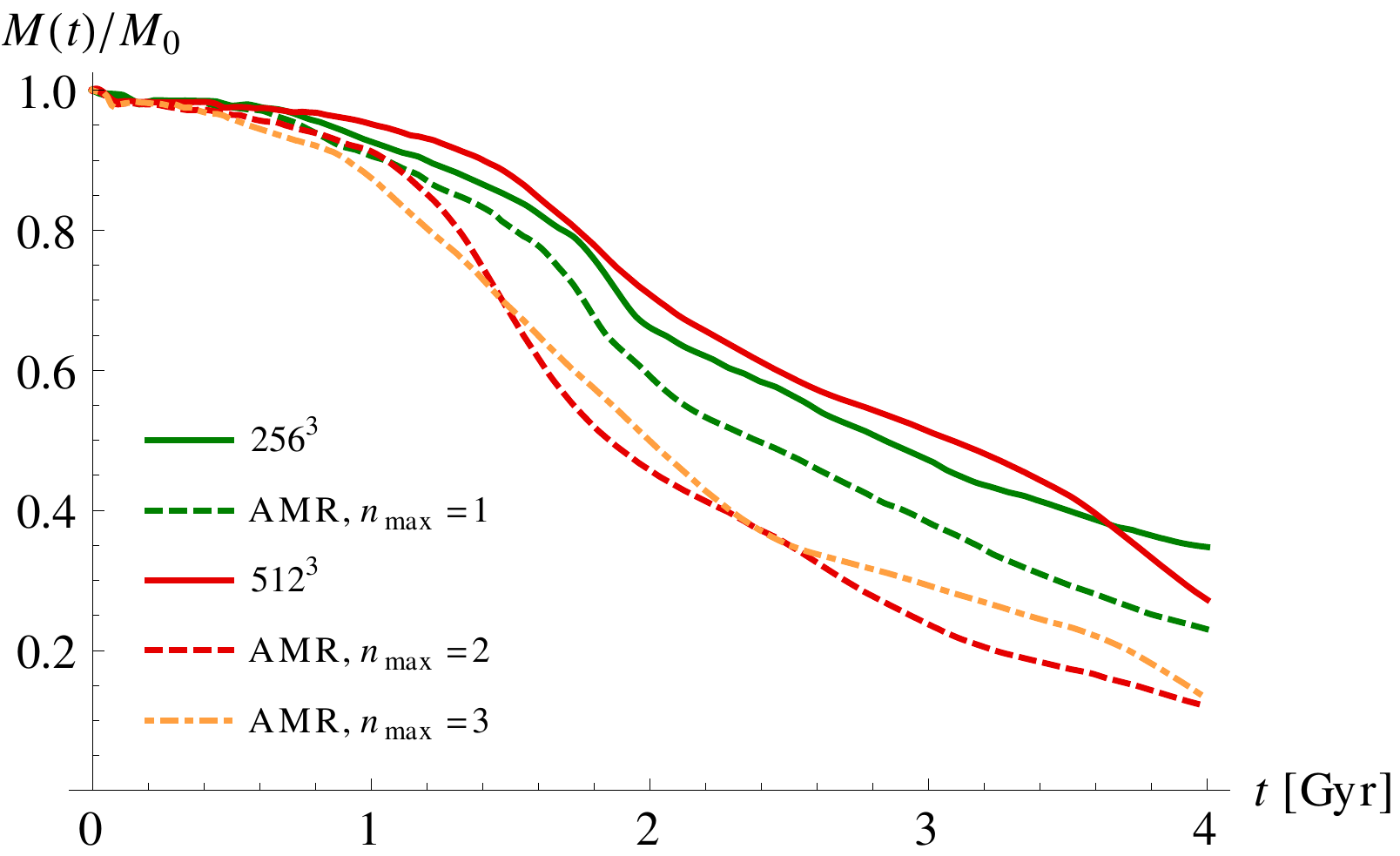}
\caption{Mass of the gravitationally bound cloud as function of time in AMR (dashed and dot-dashed lines) and uniform-grid (solid lines)
	simulations with the resolutions listed in Table~\ref{tb:runs} for {\sc Nyx} 
	(left) and {\sc Enzo} (right). Simulations with the same effective resolutions are shown in the same colour.}
\label{fig:mass_grav}
\end{figure*}

In Fig.~\ref{fig:mass_grav}, we compare the mass stripping in the different simulations performed with {\sc Nyx} and {\sc Enzo} 
(see Table~\ref{tb:runs}).
To define the cloud interior, we set $f=2$ in equation~(\ref{eq:entropy}) because the entropy contrast between the cloud center 
and the background is smaller by about a factor of $4$ compared to the case without gravitational potential. 
Again we observe systematic differences between the two codes. 
Generally, the mass stripping progresses faster in the {\sc Nyx} runs, with a steep decline from $t\approx 0.5\,\mathrm{Gyr}$ 
onwards. In this regime, instabilities at the cloud-background interface enter the non-linear regime and vortex shedding causes
the ablation of relatively large chunks from the cloud (see top panels in Fig.~\ref{fig:grav_nyx_amr3x2}).
The mass-stripping levels off after about $2\,\mathrm{Gyr}$. The frontal surface of the cloud has become more or less smooth
at this point and only small vortices are produced in a belt around the cloud (see middle panels in Fig.~\ref{fig:grav_nyx_amr3x2}).
For {\sc Enzo}, we find a qualitatively similar behavior (right plot in Fig.~\ref{fig:mass_grav}), 
but the transition between the initial stripping and the turbulent wake regime is much more gradual. Similar to the
simulations of the unbound cloud, the mass stripping progresses significantly slower compared
to {\sc Nyx}, particularly in the uniform-grid runs. This points to a genuine influence of the hydro solver. 
Moreover, the {\sc Enzo} AMR runs show strong deviations from the corresponding uniform-grid runs with the same effective resolution. For {\sc Nyx}, on the other hand, the mass stripping is nearly independent of numerical resolution and adaptivity. 
We further elaborate on these differences below.

\begin{figure*}
\centering
  \includegraphics[width=\linewidth]{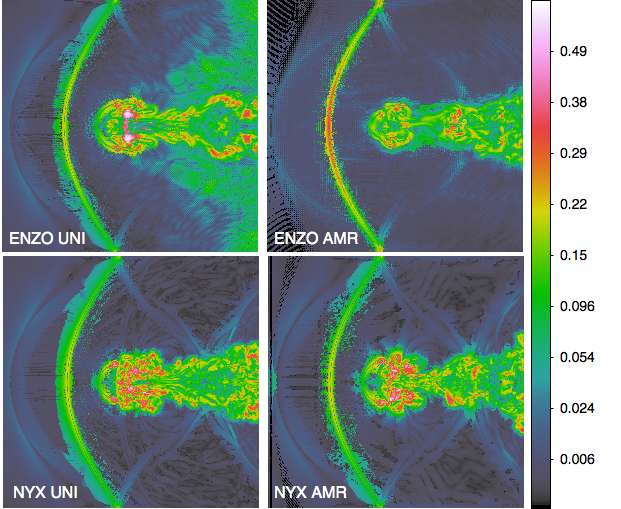}
\caption{Maps of turbulent velocity modulus normalized by the upstream wind velocity $v_{\rm b}$ at $t=3\;\mathrm{Gyr}$ for 
 	simulations of the subcluster on a $512^3$ grid (left) and a $128^3$ root grid with 2 refinement levels (right). }
\label{fig:msc_turbo}
\end{figure*}

As in the previous Section, we complement the mass statistics with an analysis of the turbulent velocity field
by applying the multi-scale filtering method. A comparison of the turbulent velocities at $t=3\;\mathrm{Gyr}$ 
for an evolutionary stage in the the slow mass stripping phase
highlights the differences between {\sc Nyx} and {\sc Enzo}, which are even more pronounced than in the case of the
unbound cloud. For the $512^3$ uniform grid (left panels in Fig.~\ref{fig:msc_turbo}), the flow
has a rather high degree of symmetry in the {\sc Enzo} run, while the subcluster
appears to be more turbulent in the {\sc Nyx} run. Moreover, {\sc Enzo} produces a noisy
pattern similar to what can be seen in Fig.~\ref{fig:filter1} for the unbound cloud. 
One can also see that fluctuations are seeded when the wind enters the gravitational well.
This spurious effect cannot be avoided because the circular boundary at radius $r_{\rm max}$ 
(see Section~\ref{sc:methods}) cannot be mapped exactly to a Cartesian grid. 
As expected, directional splitting produces
stronger fluctuations, particularly if AMR is used (top panels in Fig.~\ref{fig:msc_turbo}). 
A conspicuous feature in the AMR run performed with {\sc Enzo} is the strong fluctuations at the shock. 
It is likely that the noise induced by the gravitational cutoff upstream of the shock is amplified when it
passes through the shock. This effect might contribute to the significantly stronger mass stripping from the cloud.
For {\sc Nyx}, on the other hand, the flow structure is similar in both runs (bottom panels in Fig.~\ref{fig:msc_turbo}). 

\begin{figure}
\centering
  \includegraphics[width=0.5\linewidth]{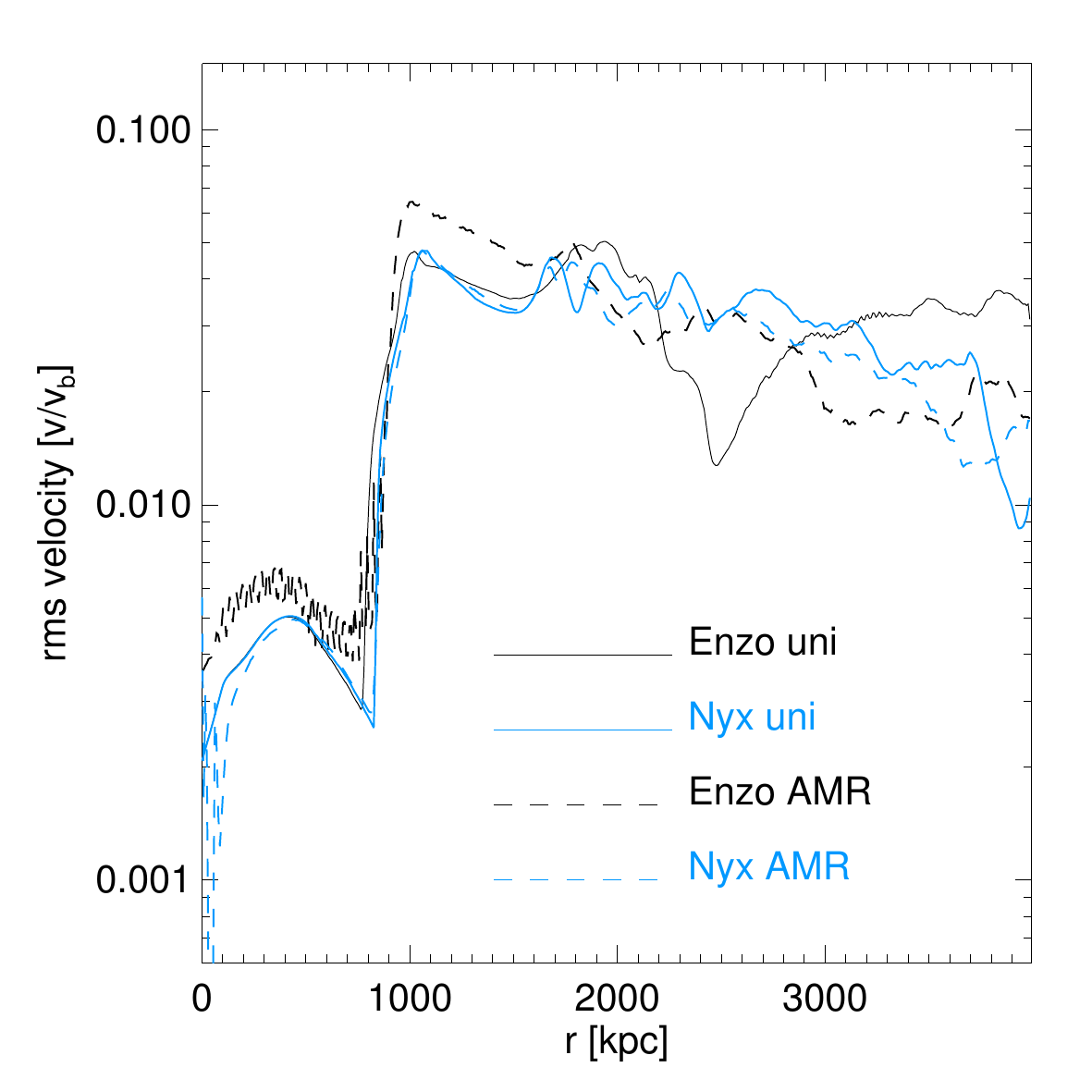}
\caption{Profiles of the root mean square velocities within slices perpendicular to the wind direction for the same datasets as
	in Figure~\ref{fig:msc_sf}. The velocities are normalized to the upstream wind velocity $v_{\rm b}$. }
\label{fig:msc_prof}
\end{figure}

Figure \ref{fig:msc_prof} shows the corresponding profiles of the rms velocity within planes perpendicular to the propagation axis of the wind. 
The profiles are quite similar in the upstream region, yet we see a small excess of pre-shock velocity
fluctuations and a larger velocity dispersion at the shock edge around $x=1000\;\mathrm{kpc}$) in the {\sc Enzo} AMR run.
Moreover, the rms velocity is significantly larger between the bow shock and the cloud in this case, which is in agreement with the turbulent velocity maps shown in Fig.~\ref{fig:msc_turbo}.
Downstream of the subcluster core (initially centered at $x = 1600\;\mathrm{kpc}$), the profiles are roughly comparable. However, the different flow structures in the {\sc Enzo} and {\sc Nyx} uniform-grid runs are reflected by differences in the turbulent velocity profiles.

\begin{figure}
\centering
  \includegraphics[width=0.5\linewidth]{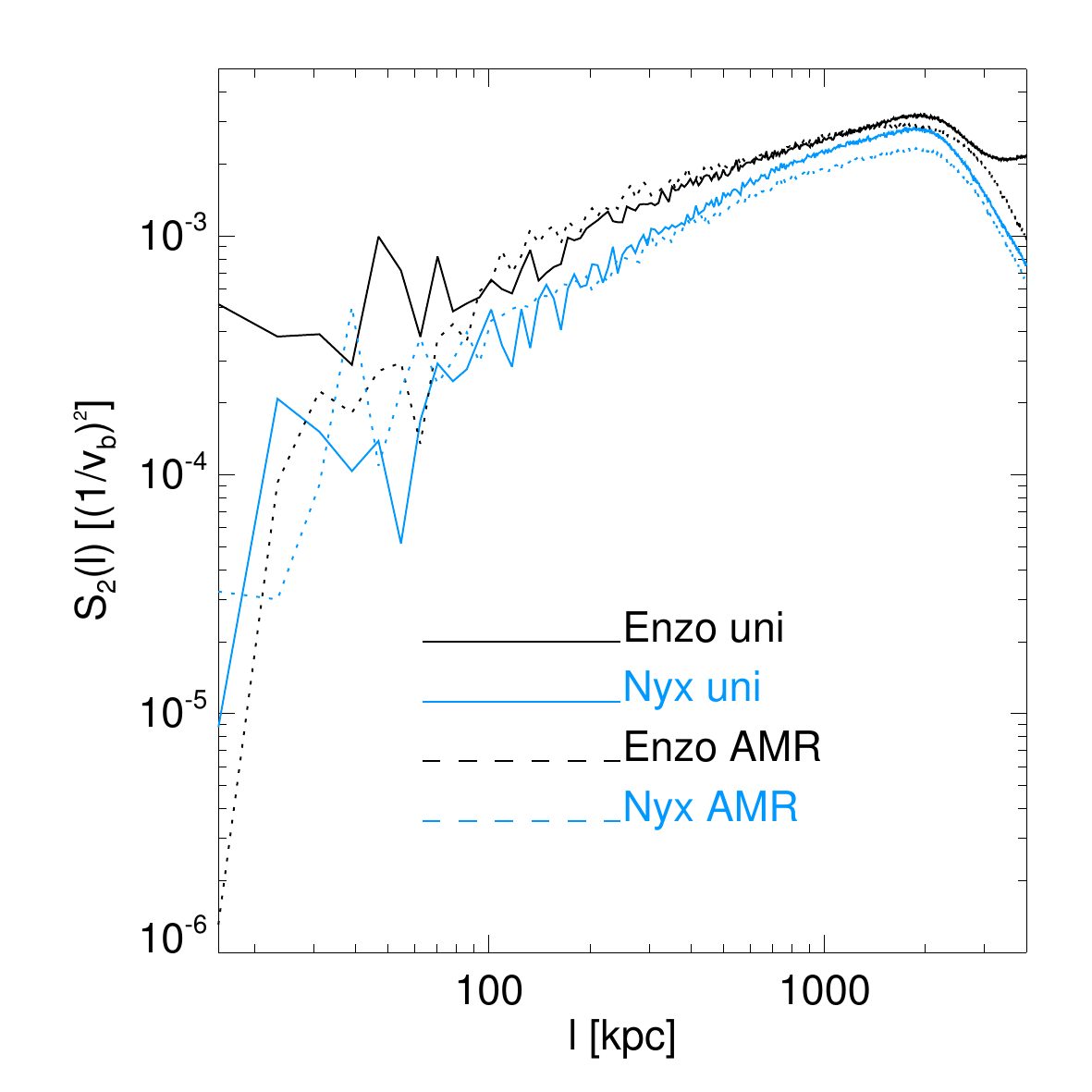}
\caption{Second-order structure function for the {\sc Enzo} and {\sc Nyx} for the same runs as in Figs.~\ref{fig:msc_turbo} 
	and~\ref{fig:msc_prof}. The velocities are normalized by the upstream wind velocity $v_{\rm b}$.}
\label{fig:msc_sf}
\end{figure}

The second-order structure functions plotted in Fig.~\ref{fig:msc_sf} are peaked at length scales much greater than the core
radius $r_{\rm core}=250\,\mathrm{kpc}$ of the subcluster. This indicates coherent structures that extend 
over the length of the tail of the subcluster, which is about $2000\;\mathrm{kpc}$. These structures contribute more to
$S_2(l)$ in the case of the $512^3$ {\sc Enzo} run than in the other runs. The structure functions are less stiff for both
{\sc Enzo} runs, which indicates an excess of small-scale structure, as suggested by the turbulent velocity maps
in Fig.~\ref{fig:msc_turbo}. To a lesser degree, differences can also be seen when comparing the two {\sc Nyx} runs:
AMR produces larger velocity fluctuations on small-scales, which is probably caused by spurious fluctuations
due to grid refinement. Interestingly, the fluctuations on length scales larger than the core radius are
slightly reduced. This suggests an enhanced decorrelation in the AMR run. 
In the case of {\sc Enzo}, these differences are much more pronounced, which
might explain the significantly stronger mass stripping and, as a consequence, the faster decrease of the size of the subcluster
core compared to the constant-resolution case. 

\section{Conclusions}
\label{sc:concl}

We performed numerical simulations of a simple cloud-in-a-wind problem with the cosmological AMR codes
{\sc Enzo} and {\sc Nyx}. The setup follows \cite{IapiAda08}, who considered an idealized model for
the minor merger of subcluster into the ICM of a big cluster. In this case, a static gravitational
potential is applied to mimic the dark matter halo of the subcluster. As a variant of this problem,
we also computed the evolution of an isothermal cloud without gravitational potential in an elongated box,
similar to the scenario investigated by \cite{AgerMoore07}.

\begin{figure*}
\centering
  \includegraphics[width=0.48\linewidth]{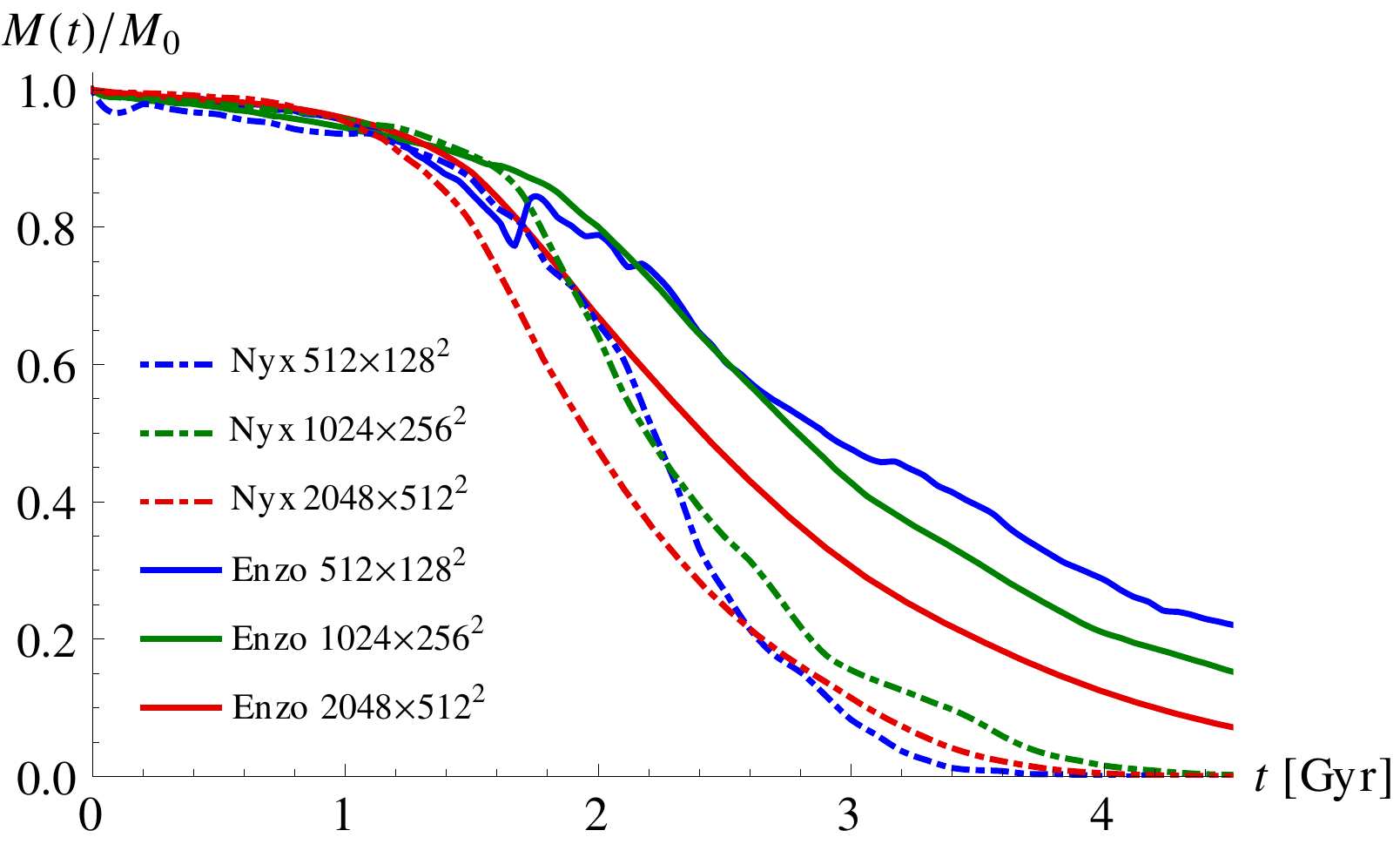}
  \includegraphics[width=0.48\linewidth]{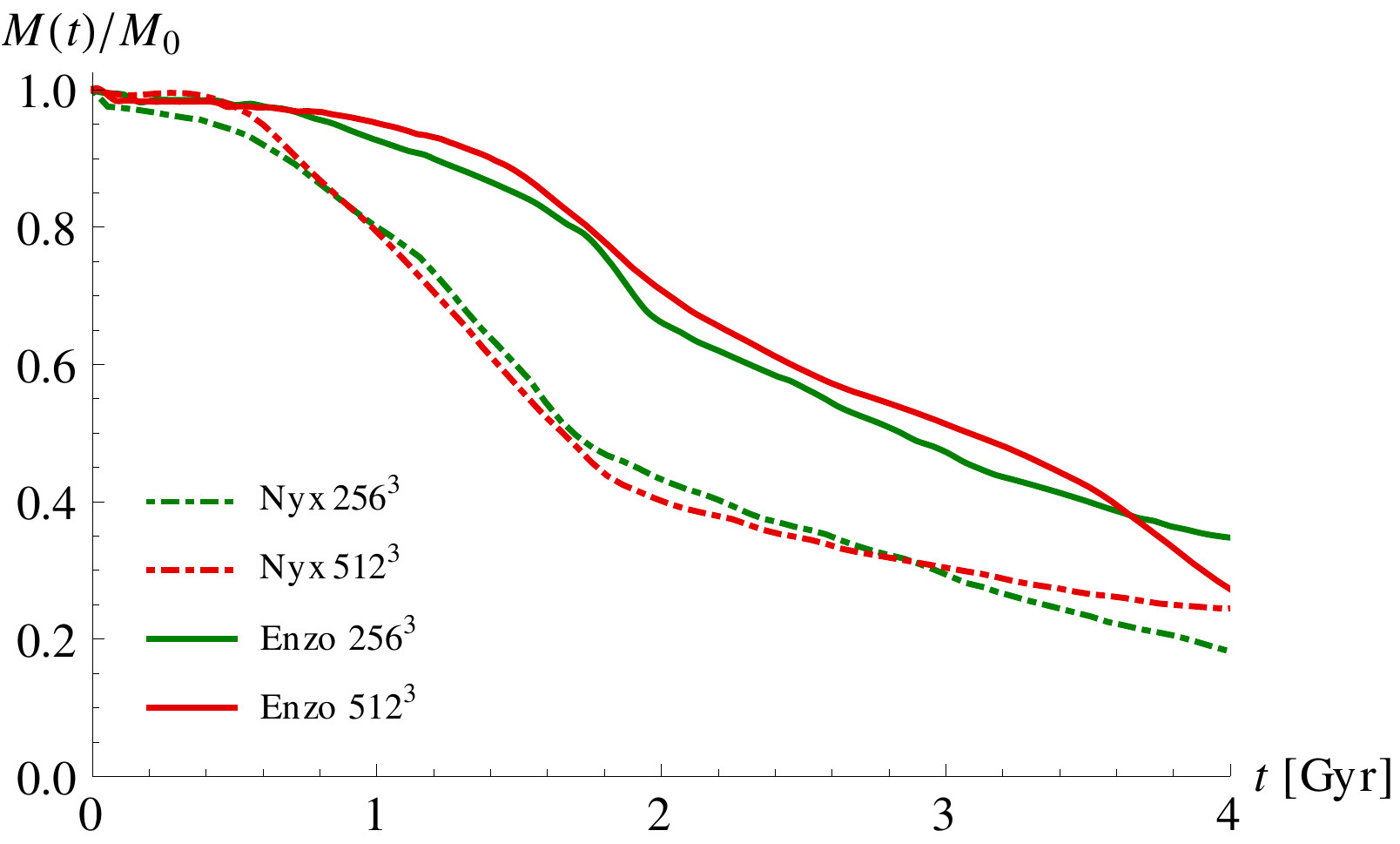}
\caption{Comparisons of the mass stripping from the unbound cloud (left) and the subcluster (right) 
	for the uniform-grid simulations with {\sc Enzo} (solid lines) and {\sc Nyx} (dot-dashed lines).
	Equivalent resolutions are shown in the same colors.}
\label{fig:mass_uniform}
\end{figure*}

As in \cite{IapiAda08} and \cite{AgerMoore07}, we computed the mass stripping from the cloud as 
basic statistical diagnostics of the effect of the wind on the cloud. An important difference to the
previous studies is that we define the cloud interior by an entropy threshold rather than a lower
bound on the density and an upper bound on the temperature. This leads to a smoother and almost
monotonic evolution of the cloud mass. As shown in the plots in Fig.~\ref{fig:mass_uniform},
the mass stripping reveals systematic differences between the
codes for both the unbound and bound clouds even if numerical grids with uniform resolution are used. 
The sensitivity to resolution is generally stronger in the case of the unbound cloud (left plot), which
is probably a consequence of the rapid destruction of the compact cloud by shear. Although the deviation
between the two codes tends to become less for increasing resolution, the deviations 
are more pronounced than the differences between runs with given resolution. This trend can be seen
very clearly for the subcluster (right plot), where the sensitivity to resolution is quite low
for uniform-grid runs, but the evolution of the cloud mass differs substantially for {\sc Enzo} and {\sc Nyx}.
The impact of the employed  hydro solvers on the flow evolution is confirmed by analyzing statistical properties 
of the turbulent flow  by means of the multi-scale filtering technique of \citet{VazzaRoed12}. 
While the power spectra of turbulence on intermediate scales and
the profiles of the RMS velocity fluctuation are generally in good agreement, 
we find deviations on scales of the order of the grid scale, which appear to be related to 
pronounced fluctuations in downstream regions in the case of {\sc Enzo}.

The various discrepancies between the two codes we found in our study might stem from the following factors:

\begin{itemize}
\item\emph{Adaptive mesh refinement}: Fine-coarse boundaries inevitably introduce some numerical noise. While the
	flow realizations clearly diverge when AMR simulations are compared to uniform-grid simulations, which
	is simply a consequence of non-linear dynamics, statistical properties are quite robust, at least for
	{\sc Nyx}. In the case of the subcluster, we find large deviations between the AMR and uniform-grid runs
	performed with {\sc Enzo}. Since this is not observed for the gravitationally 
	unbound cloud, deviations might be caused by other sources of numerical noise, for example, the
	artificial cutoff of the gravitational potential, which is specific to the subcluster setup.
	The multi-scale filtering analysis suggests that fluctuations induced at cutoff radius are amplified
	by refinement in the vicinity of the bow shock, which in turn causes stronger mass stripping. 
	In the {\sc Nyx} AMR simulations, this effect appears to be much weaker. 
\item\emph{Directional splitting}: Contrary to what one might expect, the unsplit solver implemented in {\sc Nyx}
	tends to produce more asymmetric structures and faster mass stripping compared to the solver with directional
	splitting in {\sc Enzo}. This can be understood as a consequence of the stronger alignment effects with the grid
	if directional splitting is used. Apart from that, also different flux limiters can play an
	important role for the evolution of instabilities, as demonstrated by~\citet{AlmBeck10} for the Rayleigh-Taylor instability.
	The evolution of the subcluster is strikingly different in
	{\sc Nyx} and {\sc Enzo} simulations: While a fully turbulent wake is obtained with {\sc Nyx},
	vortices tend to be more coherent in {\sc Enzo} simulations, particularly if the grid is uniform. To a certain degree,
	AMR breaks up this structures, which might contribute to the discrepancies between the
	uniform-grid and AMR simulations performed with {\sc Enzo}. 
\item\emph{Low Mach numbers}: Although the wind has supersonic speed, the vortices shedded from the cloud and the
	turbulent motions in the wake have low Mach numbers ($\lesssim 0.1$). 
	Standard PPM, as implemented in {\sc Enzo}, does not optimally perform in this regime 
	\citep{GuiMurr04}. 
	The hydro solver implemented in {\sc Nyx} was especially tuned to deal with hypersonic flow in cosmological simulations and, 
	consequently, cannot be expected to
	perform much better than the solver in {\sc Enzo} 
	(see, for example, \citealt{NonaAlm10} for the treatment of low-Mach flows). 
	Nevertheless, {\sc Nyx} appears to avoid some of the numerical artifacts produced by the PPM implementation in {\sc Enzo}. 
	Since the ICM is very hot, low-Mach flow is certainly encountered in simulations of cosmological structure formation
	\citep[e.~g.][]{LauKravt09}.
\end{itemize}

The astrophysical implications found by \cite{IapiAda08}, namely that minor mergers contribute to turbulence production
in the ICM, are confirmed by our simulations. 
The development of instabilities and gas stripping has recently been observed for minor mergers by \citet{RussFab14} and 
\cite{EckMol14}.
As demonstrated by \citet{RoedKrII}, however, the viscosity of the ICM could significantly influence the 
instabilities and the turbulent wake.

As far as numerics is concerned, 
the lesson to be taken from our study is that even for simple, purely hydrodynamical models, significantly different solutions
can result from numerical simulations if non-linear dynamics is involved. Even worse, our results for the mass stripping
from the cloud demonstrate that different codes can ``converge" toward different solutions with increasing numerical resolution.
Consequently, the notion that numerical flow representations are, in a statistical sense, close to the exact solution if changes between one resolution level and the next become sufficiently small should be taken with a large grain of salt. 

Despite its apparent simplicity, the cloud in a wind is a particularly tough problem for compressible hydro codes because it
involves symmetries in the initial conditions, which are broken by numerics, a transition from laminar to turbulent flow,
and a combination of supersonic wind with low-Mach turbulent velocity fluctuations. Although some of these difficulties are 
also encountered in full simulations of cosmological structure formation, the randomness of large-scale structure presumably 
reduces the impact of numerics. Nevertheless, it is important to further advance numerical methods and to regularly compare codes for complex astrophysical applications. 
This should complement the matching of more realistic simulations with complicated physics to astrophysical observations, which is only possible through many layers of modeling, data processing, 
and interpretation. 

{\sc Enzo} and {\sc Nyx} pass standard hydro tests with known analytic solutions 
very well, yet show divergent behavior in a multi-dimensional non-linear flow.
Such problems cannot be integrated into the now commonly used regression testing because they require 
non-negligible computational resources and their analysis is non-trivial. 
In particular, since we do not know the exact solutions, we can only resort to statistical comparisons. 
Although the results of our study reveal uncertainties that are inherent to numerical simulations, 
it turns out to be very difficult to disentangle the various factors discussed above. 
More systematic comparisons could be supported by code developers by providing more congruent sets of switches
to select different solvers, flux limiters, fallbacks, and reconstruction.
Furthermore, basic instances of non-linear hydrodynamics such as the Rayleigh-Taylor instability 
or simple planar shear layers, for which the growth rate of the Kelvin-Helmholtz instability can be computed in the linear 
and non-linear regimes, should be analyzed.\footnote{For example, \citet{JunkWal10} and \citet{NalLyr12} 
	solve two-dimensional Kelvin-Helmholtz 	instability test problems with several codes, including {\sc Enzo}. 
	However, to investigate effects occurring in turbulent flows, 
	three-dimensional simulations are necessary because the non-linear dynamics of vortices is fundamentally different
	in two- and three-dimensional flows. While finalizing our paper, an impressive suite of test problems presented was 		released by \citet{Hop14}. Although a finite volume code is used for comparison, this study focuses on fundamentally 		different discretization methods.} Both types of instabilities are relevant
for the subcluster in a wind and might shed further light on the root of the differences in the numerical solutions found 
in this work.

\section*{Acknowledgments}

We thank the referees, whose suggestions turned this article into substantially more elaborate study
than what we had originally in mind.
We are grateful to Peter Nugent and others in the Computational Cosmology Center and the Center
for Computational Sciences and Engineering for supporting the development of {\sc Nyx}. 
The Enzo code is the product of a collaborative effort of scientists at many universities and US national laboratories.
Moreover, we acknowledge the yt toolkit by \citet{TurkSmith11} that was used for the analysis and visualization of the data.  
We thank Jens Niemeyer for initiating this study. Our simulations and postprocessing were performed with the 
HLRN II facilities in Hannover under project ID nip00020 and with SuperMUC at the Leibniz Supercomputing Centre 
in Garching under project pr95he. 
The work at LBNL was supported by the SciDAC Program and the Applied Mathematics Program of the 
U.S. Department of Energy under Contract No. DE-AC02-05CH11231. 
F.~V. acknowledges the computational resources at the Juelich Supercomputing Centre (JSC), under project HH222, and support 
from grant FOR1254 from the Deutsche Forschungsgemeinschaft.

\bibliographystyle{model2-names}
\bibliography{msc.bib}







\end{document}